\documentclass[12pt]{article}
\usepackage{mheck}

\def\bphi{\boldsymbol{\Phi}}
\def\bfT{\boldsymbol{T}}
\def\bfL{\boldsymbol{L}}

\date{July, 2015}

\institution{SISSA}{Scuola Internazionale Superiore di Studi Avanzati, via Bonomea 265,  34100 Trieste, ITALY}
\institution{harvard}{Jefferson Physical Laboratory, Harvard University, Cambridge, MA 02138, USA}

\title{Higher $S$--dualities and\\
Shephard--Todd groups}

\authors{Sergio Cecotti \worksat{\SISSA} \footnote{e-mail: {\tt cecotti@sissa.it}} and 
Michele Del Zotto \worksat{\harvard} \footnote{e-mail: {\tt eledelz@gmail.com}}}

\abstract{Seiberg and Witten have shown that in $\cn=2$ SQCD with $N_f=2N_c=4$
the $S$--duality group $PSL(2,\Z)$ acts on the flavor charges, which are weights of Spin(8), by triality. There are other $\cn=2$ SCFTs in which $SU(2)$ SYM is coupled to strongly--interacting non--Lagrangian matter: their matter charges are weights of $E_6$, $E_7$ and $E_8$ instead of Spin(8). The $S$--duality group $PSL(2,\Z)$ acts on these weights: \textit{what replaces Spin(8) triality for the $E_6,E_7,E_8$ root lattices?} 

In this paper we answer the question. The action on the matter charges of (a finite central extension of) $PSL(2,\Z)$  factorizes trough the action of the exceptional Shephard--Todd groups $G_4$ and $G_8$ which should be seen as complex
analogs of the usual triality group $\mathfrak{S}_3\simeq \mathrm{Weyl}(A_2)$.  
 Our analysis is based on the identification of  $S$--duality for $SU(2)$ gauge SCFTs with the group of automorphisms of the cluster category of weighted projective lines of tubular type.
}

\begin{document}

\maketitle

\tableofcontents
\newpage



\section{Introduction and Summary}\label{intt}

We have a complete classification of the 4d
$\cn=2$ gauge theories where $SU(2)$ SYM is coupled to vector--less matter (possibly non--Lagrangian) \cite{CV11}. By \emph{vector--less} matter we mean $\cn=2$ QFTs whose BPS spectra, in all chambers, consist only of hypermultiplets: the ones having a gaugeable  $SU(2)$ symmetry are precisely the Argyres--Douglas (AD) theories of type $D_p$ ($p\geq 2$) \cite{CV11,eguci}  quark doublets being the case $p=2$.
From the classification we learn that the SCFTs in this class are in one--to--one correspondence with the orbifolds of an elliptic curve $\ce$ i.e.
\begin{equation}\label{orbifold}\ce/\Z_p\quad \text{where }p=1,2,3,4,6.\end{equation}
For $p>2$ the curve $\ce$ should have complex multiplication by the appropriate quadratic field namely\footnote{ $\omega$ is a primitive third root of unity, i.e.\! a solution to the cyclotomic equation $\omega^2+\omega+1=0$.} $\mathbb{Q}(\omega)$ for $p=3,6$ and $\mathbb{Q}(i)$ for $p=4$. This observation plays a crucial role below.

We have five SCFTs in this class of $SU(2)$ gauge theories. The SCFT is a Lagrangian model iff the  modulus $\tau$ of $\ce$ is a free parameter, i.e.\! for $p=1,2$ which correspond, respectively, to $\cn=2^*$ and SQCD with $N_f=4$. $p=3,4,6$ yield three additional non--Lagrangian ($\equiv$ intrinsically strongly coupled) SCFTs. From the viewpoint of \cite{CV11} it is more natural to state this classification as a one--to--one correspondence between this class of SCFTs and the \emph{star} graphs (possibly with multiple edges\footnote{ Non simply--laced star graphs corresponds to matter in $SU(2)$ representations of isospin $>1/2$.}) which are affine Dynkin diagrams. There are five such affine stars
\begin{equation}\label{affine}
\mathfrak{g}^{(1)}=A^{(1)}_1,\ D_4^{(1)},\ E_6^{(1)},\ E_7^{(1)},\ E_8^{(1)},
\end{equation}
which correspond to the five orbifolds \eqref{orbifold}. In the first two models, the Lagrangian ones, 
$\mathfrak{g}^{(1)}$ is also the affinization of the flavor symmetry algebra $\mathfrak{g}$ which is, respectively, $\mathfrak{su}(2)$ and $\mathfrak{so}(8)$: thus in the Lagrangian models the flavor charges are weights of $\mathfrak{g}$. In the non--Lagrangian theories the matter consists of strongly interacting systems with their own conserved 
electric and magnetic charges in addition to the flavor ones. Although 
the Dirac pairing between the internal charges of the matter is no longer trivial, yet it remains true that the matter charges\footnote{ Throughout this paper, by \emph{matter charges} we mean all conserved charges of the $\cn=2$ QFT but the $SU(2)$ electric and magnetic ones.} take value in the weight lattice of the corresponding finite--dimensional Lie algebra $\mathfrak{g}$ which, for the non--Lagrangian models, is $E_6$, $E_7$, or $E_8$ (see \S.\,\ref{overview}).

These theories, already described in \cite{CV11,cattoy,CDZG}, recently have been constructed also as toroidal compactifications of certain 6d $(1,0)$ SCFTs \cite{DZXV}.
From the internal torus, all five
(mass deformed) 4d SCFTs inherit
a $PSL(2,\Z)$ group of $S$--dualities.
In the $p=2$ case Seiberg and Witten \cite{SW2} have shown that $PSL(2,\Z)$ acts on the flavor charges by $SO(8)$  triality; the triality group $\mathfrak{S}_3$ being identified with the modular quotient $PSL(2,\Z)/\Gamma(2)$.
By the same token, for $p=3,4,6$ we have a non--trivial action of $S$--duality, hence of the modular group $PSL(2,\Z)$,
on the lattice of the matter charges, i.e.\! on the weight lattices of $E_6,E_7,E_8$. This action should be thought of as a generalization of the triality action on the weights of $SO(8)$ to the weights of the exceptional Lie groups $E_6,E_7,E_8$. 

At first sight this statement seems rather odd: from Lie algebra theory we do not expect any higher rank analog of $SO(8)$ triality. Yet physics \emph{predicts} its existence. 

 The full duality group $\mathbb{S}$ is actually an extension of the modular group
 \begin{equation}
 1\to\mathscr{W}\to \mathbb{S}\to PSL(2,\Z)\to 1,
 \end{equation}
where $\mathscr{W}$ is the `obvious' group of physical symmetries acting on the lattice of conserved charges. For the Lagrangian models
$\mathscr{W}$ is simply the Weyl group of the flavor symmetry. In the general case $\mathscr{W}$ is a well--understood finite group of symmetries of the matter system (seen as decoupled from the Yang--Mills sector) which fixes the charge to be gauged.
$\mathscr{W}$ is also the kernel of the action of $\mathbb{S}$ on
the Yang--Mills electric/magnetic charges, on which only the quotient group $PSL(2,\Z)$ acts effectively. The quotient group of
$\mathbb{S}$ which acts effectively on
the matter charges, $\mathbb{S}_\mathrm{matter}$,
is the \emph{finite} group
\begin{equation}\label{qqqq}
1\to\mathscr{W}\to \mathbb{S}_\text{matter}\to PSL(2,\Z/p\,\Z)\to 1,\qquad p=1,2,3,4,6,
\end{equation}
whose action on the root lattice of $\mathfrak{g}$ preserves the Cartan inner product as well as the Dirac skew--symmetric pairing. For $p=1,2$ eqn.\eqref{qqqq} was obtained by Seiberg and Witten \cite{SW2}. Its extension to the non--Lagrangian cases looks rather natural, a simple `analytic continuation in $p$'.

Since the action of $\mathscr{W}$ on the matter charges is obvious, to understand the `higher versions of triality' it is enough to 
understand the action of the quotient group

\begin{equation}
PSL(2,\Z/p\,\Z)\simeq PSL(2,\Z)\big/\Gamma(p)\qquad \text{for }p=2,3,4,6,
\end{equation}
where $\Gamma(p)\subset PSL(2,\Z)$ is the principal congruence subgroup of level $p$ \cite{cong}.
$p=6$ is special since\footnote{ There is a much stronger reason why $p=6$ is different. As Klein proved in 1884 \cite{klein} the group $PSL(2,\Z)/N(6)$ has \emph{infinite} order, while for $p=1,2,3,4$, $PSL(2,\Z)/N(p)=PSL(2,\Z/p\,\Z)$ (see \S.\ref{crg}).}
\begin{equation}\label{chinese}
PSL(2,\Z/6\Z)=PSL(2,\Z/2\Z)\times PSL(2,\Z/3\Z).
\end{equation}
However, it is preferable to study the action of a subgroup
$G_\text{matter}\subset \mathbb{S}_\text{matter}$ which is a central extension 
of $PSL(2,\Z/p\,\Z)$ by a finite Abelian group
of the form $(\Z/2\Z)^k$. Studying the central extension $G_\text{matter}$, rather than $PSL(2,\Z/p\,\Z)$ itself, allows to discuss all five models in an unified way \emph{via} the theory of reflection groups.

Let us explain. Weyl groups should be thought of as reflection groups defined over the rationals $\mathbb{Q}$, and Coxeter groups as real reflection groups. For $N_f=4$ SQCD ($p=2$) the action of $PSL(2,\Z/2\Z)$ is through $SO(8)$ triality which is a \emph{rational} reflection group: indeed, in concrete terms, the triality group is the quotient
\begin{equation}\mathbb{S}_\text{matter}\big/\mathrm{Weyl}(SO(8))\equiv\mathrm{Weyl}(F_4)\big/\mathrm{Weyl}(SO(8)).\end{equation}
In passing from $p=2$ to $p=3,4,6$ what we have to do is to replace reflection groups defined over $\mathbb{Q}$ with reflection groups defined over the appropriate
complex multiplication fields $\mathbb{Q}(\omega)$ or $\mathbb{Q}(i)$. Roughly speaking, the abstract form of $S$--duality is the same for $p>2$ as for the SQCD $p=2$ model  (where it is given by $SO(8)$ triality) but structures that
 in the $p=2$ case are defined over the ground field $\mathbb{Q}$ get replaced by structures defined over the complex quadratic fields
 $\mathbb{Q}(\omega)$ and $\mathbb{Q}(i)$.

 Reflection groups defined over such quadratic fields are special instances of \emph{complex reflection groups}. The \emph{finite}
 complex reflection groups have been fully classified by Shephard and Todd \cite{ST}. From their classification we read the complete list of
the  ``higher triality'' groups $G_\mathrm{matter}$ which act on the matter charges of the $p=3,4,6$ models, i.e.\! on the weight lattices $\Gamma^\mathrm{w}_\mathfrak{g}$ of 
 $E_6$, $E_7$ and $E_8$. The classification also yields the decomposition of the vector space $\Gamma^\mathrm{w}_\mathfrak{g}\otimes \C$ in irreducible representations of  $G_\mathrm{matter}$, and thus completely specifies how $PSL(2,\Z/p\,\Z)$, and hence the full modular group $PSL(2,\Z)$, acts on the matter charges.

\begin{table}
\begin{minipage}{460pt}
\begin{center}
\caption{\label{comgroups}The relevant complex reflection groups}
\begin{tabular}{c|c|c|c|c|c}\hline\hline
field $\mathbb{F}$ & $\phantom{\bigg|}{\text{reflection}\atop \text{group}}$ & ${\text{abstract}\atop \text{group}}$ & ${\text{extension of a quotient}\atop \text{of the modular group}}$& ${\text{reflection}\atop
\text{group graph}\footnote{ Usually the 2's in the nodes of the $A_2$ Dynkin graph are omitted.}}$ & ${\text{McKay affine}\atop
\text{graph}\footnote{ \emph{More precisely}: the McKay graph of the $SU(2)$ subgroup which has the same image in $PSU(2)$.}}$\\\hline
$\mathbb{Q}$ & $\phantom{\bigg|}\mathrm{Weyl}(A_2)$ & $\mathfrak{S}_3$
& $PSL(2,\Z/2\Z)$ & \begin{tiny}$\xymatrix{*++[o][F-]{2}\ar@{-}[r]&
*++[o][F-]{2}}$\end{tiny} & $D_5^{(1)}$ \\
$\mathbb{Q}(\omega)$ & $\phantom{\bigg|}G_4$ & ${\text{binary}\atop \text{tetrahedral}}$ & $SL(2,\Z/3\Z)$ 
 & \begin{tiny}$\xymatrix{*++[o][F-]{3}\ar@{-}[r]&
*++[o][F-]{3}}$\end{tiny}& $E_6^{(1)}$\\
$\mathbb{Q}(i)$ & $\phantom{\bigg|}G_8$ & $\Z/2\Z\ltimes \!\!\left({\text{binary}\atop \text{octahedral}}\right)$ & $\Z/2\Z\ltimes\! SL(2,\Z/4\Z)$ 
 & \begin{tiny}$\xymatrix{*++[o][F-]{4}\ar@{-}[r]&
*++[o][F-]{4}}$\end{tiny}& $E_7^{(1)}$\\
\hline\hline
\end{tabular}
\end{center}
\end{minipage}
\end{table}

 The relevant reflection groups for the three
 complex multiplication fields $\mathbb{Q}$, $\mathbb{Q}(\omega)$ and $\mathbb{Q}(i)$ are listed in table \ref{comgroups}. In the second column we write the group, seen as a concrete reflection group acting on a two--dimensional space, in the Shephard--Todd notation.
 In the third column we write the standard name of the corresponding abstract group. 
 In the fourth column we describe the same 
 group seen as a quotient of a central extension of the modular group $PSL(2,\Z)$; this column specifies how the action of $S$--duality group $\mathbb{S}$ on $\Gamma^\mathrm{w}_\mathfrak{g}$ factorizes through a representation of the reflection group $G_\text{matter}$. In the fifth column we draw the  graph of the reflection group \cite{Rgraph}: we stress that, in all three cases, it is the $A_2$ Dynkin graph but with order $p$ at the nodes. The uniformity of the graph expresses our rough idea that the structure of $S$--duality is independent of $p$ up to a change of the ground field. In particular, all three groups are realized as a concrete group of reflections by faithful \emph{two}--dimensional unitary representations which we denote as $W$, $R$ and $F$, respectively (see \S.\ref{crg} for full details). In the last column of the table we recall the affine Dynkin graph which is related to the given reflection group by the McKay correspondence \cite{mckay}; more precisely, the affine graph shown in the table is the one associated to the finite $SU(2)$ subgroup which has the same image in $PSU(2)$ as the reflection group $G_\text{matter}$ viewed as a subgroup of $U(2)$ \emph{via} its defining
 two--dimensional representation. Note that to the $E_6^{(1,1)}$ and
 $E_7^{(1,1)}$ SCFTs there correspond, respectively, the McKay graphs $E_6^{(1)}$ and
 $E_7^{(1)}$.

Summarizing, the reflection groups $G_\text{matter}$ acting on the matter charges, which take values in the weight lattices $\Gamma_\mathfrak{g}^\mathrm{w}$, are (see \S.\ref{crg} for the definition of the groups $G_4$, $G_8$)
\begin{equation}\label{mattergroup}
G_\text{matter}=\left\{\begin{array}{ll}
 \mathfrak{S}_3 &p=2\\
G_4&p=3 \\
G_8&p=4  \\
 \mathfrak{S}_3\times G_4&p=6.
\end{array}\right.
\end{equation}
For $p=2$ the symmetric group $\mathfrak{S}_3$ acts on the weights of $D_4$ by triality.
In this paper we describe the corresponding action of $G_\text{matter}$ for $p=3,4,6$.
We do so in two ways. First we list the irreducible representations of the matter group in eqn.\eqref{mattergroup} acting on the root vector space
$\Gamma_{E_r}\otimes \C$
\begin{align}\label{res1}
&p=3 && \Gamma_{E_6}\otimes\C \simeq \boldsymbol{1}\oplus \boldsymbol{1}\oplus R\oplus \overline{R}\\
&p=4 && \Gamma_{E_7}\otimes\C \simeq \boldsymbol{1}\oplus W\oplus F\oplus \overline{F}\label{eeqqq}\\
&p=6 && \Gamma_{E_8}\otimes\C \simeq \chi\oplus \overline{\chi} \oplus (W,\boldsymbol{1})\oplus (\boldsymbol{1},R)\oplus (\boldsymbol{1},\overline{R}),\label{res3}
\end{align}
where: $\boldsymbol{1}$ is the trivial representation; $W$, $R$ and $F$ are the defining $2$--dimensional representations of 
$\mathfrak{S}_3$, $G_4$ and $G_8$, respectively; $\overline{R}$, $\overline{F}$ are their conjugates ($W$ is real); $\chi$ is a one--dimensional representation, namely a primitive character of the
Abelian quotient\footnote{ From table \ref{comgroups} one has $G_4\simeq \Z/2\Z \ltimes \mathfrak{A}_4$. $V_4\simeq \Z/2\Z\times \Z/2\Z$ is the Klein \emph{Vierergruppe}
\cite{klein}, the unique non--trivial normal subgroup of the alternating group $\mathfrak{A}_4$.} 
\begin{equation}
\Z/6\Z\simeq \Z/2\Z\times \Z/3\Z\simeq\mathfrak{S}_3\big/\mathfrak{A}_3\times \mathfrak{A}_4\big/V_4\end{equation}
 In eqn.\eqref{eeqqq} by $W$ we mean the 2--dimensional representation of $\Z/2\Z\ltimes SL(2,\Z/4\Z)$ defined by the degree 2 representation $W$ of $\mathfrak{S}_3$ \emph{via} the canonical mod 2 surjection 
\begin{equation}\label{zzz34}\Z/2\Z\ltimes SL(2,\Z/4\Z)\longrightarrow PSL(2,\Z/2\Z)
\xrightarrow{\ W\ } GL(W).\end{equation}
The action of $G_\text{matter}$ on
$\Gamma_{E_r}\otimes\C$ preserves the lattice $\Gamma_{E_r}$, the inner product in root space given by the Cartan matrix, and the
Dirac electro--magnetic pairing.
For concreteness, in appendix \ref{explicitmatrices} we also give a very explicit realization of the action of $S$--duality on the charges in terms of integral $(r(\mathfrak{g})+2)\times(r(\mathfrak{g})+2)$ matrices acting on the full
charge lattice $\Gamma$ (which includes the matter charges as well as the Yang--Mills electric and magnetic ones).
\medskip

The rest of the paper is organized as follows. In section 2 we collect the basic tools of the homological approach to $S$--duality.
In section 3 we reconsider $\cn=2$ SQCD with two colors and four flavors as a warm--up;
here we recover the Seiberg--Witten result in two ways: first in a rather naive but very concrete approach, and then from a more intrinsic group--theoretical perspective. In section 4 we describe the
$E_6^{(1,1)}$, $E_7^{(1,1)}$ and $E_8^{(1,1)}$ cases. In appendix A we write explicit $(r(\mathfrak{g})+2)\times (r(\mathfrak{g})+2)$ matrices which represent the action of $S$--duality on the conserved charges of the theory, and list some of the beautiful identities they satisfy.  In appendix B we show how the present approach is related to the one in \cite{CV11,ACCERV} by \emph{cluster--tilting.}

\section{Homological approach to $S$--duality}\label{general}

In this section we review $S$--duality for the
four (mass--deformed) SCFTs $D^{(1,1)}_4$,
$E^{(1,1)}_6$, $E^{(1,1)}_7$ and $E^{(1,1)}_8$
following the mathematical literature \cite{GL1,lenzing1,lenzing2,lenzingmeltzer,meltzer,BKL} (see also \cite{ringelDER}).
\paragraph{Remark.} For definiteness, here we use the 4d definition of the relevant SCFTs,
see refs.\!\cite{CV11,cattoy}. Alternatively, one could have adopted the 6d viewpoint of ref.\!\cite{DZXV}, and in particular their mirror Landau--Ginzburg description of the
$\ce/\Z_p$ orbifolds. By a theorem of Orlov (\!\!\cite{orlov} \textbf{Theorem 2.5.}\textit{(iii)}, see also the discussion in \cite{ADEchain}), the category of $B$--branes for the relevant Landau--Ginzburg models is \emph{equivalent} to the derived category of coherent sheaves on the corresponding weighted projective lines of tubular type, which is the central object of our analysis, see \S.\ref{wpl}. Then all our considerations apply directly to the Landau--Ginzburg set up.
Needless to say, the equivalence of the two categories is an instance of the $2d/4d$ correspondence advocated in \cite{CNV}.

\subsection{Overview}\label{overview}

The relation between eqns.\eqref{orbifold},\eqref{affine} and the mass--deformed SCFTs is as follows\footnote{ For the justification of these assertions, see footnote \ref{giustification}. More technical details in appendix \ref{tilting}.}. The BPS particles of the QFT are given by the (quantization of) continuous families of \emph{stable} objects in a certain orbifold category $\cc(\ce/\Z_p)$ of the derived category of coherent sheaves on the orbifold $\ce/\Z_p$ or, equivalently \cite{equivalenceorbifold}, on the  orbifold $\mathbb{P}^1/\Gamma_p$ where $\Gamma_p\subset SU(2)$ is the finite subgroup associated to the affine Lie algebra $\mathfrak{g}^{(1)}$ by the McKay correspondence\footnote{ We write
$\mathsf{coh}\,X/G$ as a shorthand for the category of $G$--equivariant coherent sheaves
on $X$.} \cite{mckay} 
\begin{equation}\label{orbitcat}
\cc(\ce/\Z_p)=\mathsf{D}^b(\mathsf{coh}\,\ce/\Z_p)\big/\mathscr{C}\simeq \mathsf{D}^b(\mathsf{coh}\,\mathbb{P}^1/\Gamma_p)\big/\mathscr{C}.
\end{equation}
Here $\mathscr{C}$ is an infinite cyclic subgroup of $\mathrm{Aut}(\mathsf{D}^b\,\mathsf{coh}\,\ce/\Z_p)$ to be described later.
$\cc(\ce/\Z_p)$ is the \emph{cluster category} \cite{keller,BKL} of the orbifold $\ce/\Z_p$ whose relevance for the physics of the BPS sector stems from the
Kontsevich--Soibelman wall--crossing formula \cite{KS}, see refs.\!\cite{CNV,Ysy}. In the present context, \emph{stable} means that the lift of the object in $\mathsf{D}^b\,\mathsf{coh}\,\ce/\Z_p$ is stable in the sense of ref.\!\cite{stableZ}; the stability condition on $\mathsf{D}^b\,\mathsf{coh}\,\ce/\Z_p$ is defined by the mass deformation we consider and the chosen point in the Coulomb branch. Since all five QFTs are complete in the sense of \cite{CV11}, we have only zero-- and one--dimensional families of stable objects which yield, respectively, hypermultiplets and vector multiplets of $\cn=2$ \textsc{susy}.

\paragraph{Charge lattices.} The  lattice of conserved QFT charges, $\Gamma$,  may then be identified with the 
Grothendieck group of the associated additive categories\footnote{ The fact that the Grothendieck group is a finite--rank lattice follows from the fact that the Abelian category $\mathsf{coh}\,\ce/\Z_p$ admits a tilting object \cite{GL1,lenzing1,lenzing2}.}
\begin{equation}
\Gamma=K_0(\mathsf{D}^b\,\mathsf{coh}\,\ce/\Z_p)\simeq K_0(\mathsf{coh}\,\ce/\Z_p).
\end{equation}
Given an object $X\in \mathsf{D}^b\,\mathsf{coh}\,\ce/\Z_p$, we write $[X]\in\Gamma$ for its Grothendieck class; if $X$ is stable, $[X]$ is the charge vector of a corresponding BPS state. In all chambers the charges of BPS particles generate $\Gamma$.
On $\Gamma$ we have a bilinear form, the Euler pairing
\begin{equation}
\langle [X],[Y]\rangle_E=\sum_{k\in\Z} (-1)^k
\dim\mathrm{Hom}^\bullet(X,Y[k]),
\end{equation}
where $Y\to Y[1]$ denotes the shift equivalence in the triangle category $\mathsf{D}^b\,\mathsf{coh}\,\ce/\Z_p$.
The Dirac pairing in $\Gamma$ is just the anti--symmetric part of the Euler one
\begin{equation}
\langle [X],[Y]\rangle_\text{Dirac} =\langle [X],[Y]\rangle_E-\langle [Y],[X]\rangle_E.
\end{equation}
 The Tits form is the integral quadratic form on $\Gamma$
 \begin{equation}q([X])=\langle [X],[X]\rangle_E.\end{equation}
The symmetric bilinear form associated to the Tits quadratic form will be written $\langle \cdot,\cdot\rangle_\text{sym}$.
 
 For $X$ stable one has \cite{ACCERV,galois}
 \begin{equation}
 q([X])=2\big(1- \mathrm{MaxSpin}([X])\big),
 \end{equation}
 where $\mathrm{MaxSpin}([X])$ is the largest possible spin for a BPS particle of charge $[X]$.
Since for a complete $\cn=2$ theory \cite{CV11,galois}, the spin of the BPS particles is bounded by 1, the quadratic form $q([X])$ is positive semi--definite; its radical 
\begin{equation}
\mathrm{rad}\,q\equiv\Big\{[X]\in\Gamma\; \big|\; q([X])=0\Big\}\subset \Gamma,
\end{equation}
is a sublattice of rank $2$ which may be identified with the lattice of electric/magnetic Yang--Mills $SU(2)$ charges;
in particular, the restriction to $\mathrm{rad}\,q$ of the Dirac pairing has the form\footnote{ In the Lagrangian case, $p=2$,
the overall coefficient $2$ is interpreted as the Cartan matrix of $SU(2)$.}
\begin{equation}p\begin{pmatrix} 0 & 1 \\ -1 & 0\end{pmatrix}.\end{equation}
We define the
\emph{matter charges lattice} as $\Gamma$ modulo the sublattice of Yang--Mills charges
\begin{equation}\label{mattercharges}\Gamma_\mathrm{matter}\simeq \Gamma\big/\mathrm{rad}\,q.\end{equation}
The above categorical identification of the charges is not the conventional one in physics (but it coincides for Lagrangian QFTs). For the three non--Lagrangian models, our $SU(2)$ electric charge is a linear combination of the physical electric charge and matter ones; 
the present conventions make the group actions more transparent (in facts, they are the obvious generalization of the factor 2 difference of normalizations for the $p=1$ and $p=2$ models pointed out in \S.16 of the original Seiberg--Witten paper \cite{SW2}).
 
By general theory of integral quadratic forms \cite{ringel} (reviewed in the present context in \cite{galois}) the Tits form $q$ induces an integral quadratic form
$\overline{q}$ on $\Gamma\big/\mathrm{rad}\,q$ which is positive--definite, hence $\Z$--equivalent to the Tits form $q_\mathfrak{g}$ on the \emph{root} lattice $\Gamma_\mathfrak{g}$ of 
a Lie algebra $\mathfrak{g}$ of $ADE$ type
\begin{equation}q_\mathfrak{g}(x_i)=\frac{1}{2} C_{ij}x_ix_j,\qquad C_{ij}\ \text{the Cartan matrix of }\mathfrak{g}.\end{equation}
For our five categories $\mathsf{D}^b(\mathsf{coh}\,\ce/\Z_p)$ with $p=1,2,3,4,6$ one finds (see \S.\ref{wpl})
\begin{equation}\overline{q} \simeq q_\mathfrak{g}\ \text{hence }\Gamma_\mathrm{matter}\simeq \Gamma_\mathfrak{g}\quad\text{where }\mathfrak{g}=A_1,\ D_4,\ E_6,\ E_7,\ E_8\ \text{respectively}.\end{equation}
In the language of refs.\!\cite{CV11,galois} the two statements $\mathrm{rank\,rad}\,q=2$ and
$\overline{q}\simeq q_\mathfrak{g}$ are summarized in the fact that the mutation class of quivers with superpotentials $(Q,\cw)$ which describe the BPS sector of our
five theories is given by the elliptic Dynkin graphs $\mathfrak{g}^{(1,1)}$ of respective type
$\mathfrak{g}=A_1, D_4,E_6,E_7,E_8$ (with all triangles oriented). The \emph{elliptic} (or \emph{toroidal} \cite{moody}) Lie algebra $\mathfrak{g}^{(1,1)}$ of type $\mathfrak{g}$  is obtained by affinization of the affine Lie algebra $\mathfrak{g}^{(1)}$ of the same type; see \cite{galois} for more details.

As a basis of the matter charges we take the simple roots $\alpha_a$ of $\mathfrak{g}$.
Then the matter charges of the BPS particle corresponding to the stable object $X\in \mathsf{D}^b\,\mathsf{coh}\,\ce/\Z_p$ are
\begin{equation}f_a(X)=\langle [X],\alpha_a\rangle_\mathrm{sym}\in \Z,\qquad a=1,2,\dots, r(\mathfrak{g}).\end{equation}
In particular, for $p=1,2$ we get back that the flavor charges take values in the weight lattice of $Sp(1)$ and $SO(8)$, respectively. The same statement holds, in the present sense, for $p=3,4,6$.

\paragraph{Serre duality and Coxeter transformation.} The Abelian category
$\mathsf{coh}\,\ce/\Z_p$ satisfies Serre duality in the form\footnote{ Here $D$ stands for the usual duality over the ground field $\C$, i.e.\! $D(-)= \mathrm{Hom}(-,\C)$.}
\begin{equation}\mathrm{Ext}^1(X,Y)=D\,\mathrm{Hom}(Y,\tau X),\qquad \tau X\equiv \omega\otimes X,\end{equation}
where $\omega$ is the dualizing sheaf. 
$\tau$ is an auto--equivalence of 
$\mathsf{coh}\,\ce/\Z_p$, and hence
of the derived category $\mathsf{D}^b\,\mathsf{coh}\,\ce/\Z_p$;
$\tau$ also plays the role of Auslander--Reiten translation \cite{book1}. Given that $\tau$
is an auto--equivalence, $\mathsf{coh}\,\ce/\Z_p$ has no non--zero injectives or projectives and the Abelian category $\mathsf{coh}\,\ce/\Z_p$ is \emph{hereditary} (global dimension $1$).

Since $\omega^p\simeq \co$, one has $\tau^p=\mathrm{Id}$. The Coxeter transformation $\boldsymbol{\Phi}\colon \Gamma\to \Gamma$ is defined by
\begin{equation}[\tau X]=\boldsymbol{\Phi}\cdot[X].\end{equation}
In particular, for all $[X],[Y]\in\Gamma$ we have
\begin{align}\label{A1}\langle \bphi\cdot[X],\bphi\cdot[Y]\rangle_E&=\langle [X],[Y]\rangle_E,\\
\langle [X],\bphi\cdot[Y]\rangle_E&=-\langle [Y],[X]\rangle_E,\label{A2}\end{align}
which implies
\begin{equation}[X]\in\mathrm{rad}\,q\quad\Longleftrightarrow\quad \bphi\cdot[X]=[X],\end{equation}
i.e.\! the Yang--Mills magnetic/electric charges are the $(+1)$--eigenvectors of $\bphi$.
Likewise, the flavor charges are the 
$(-1)$--eigenvectors  of $\bphi$. The eigenvectors associated to eigenvalues $\lambda\not=\pm1$ correspond to internal electric/magnetic charges of the matter AD systems. Note that $\bphi^p=1$, in fact (for $p>1$)
\begin{equation}\det[z-\bphi]=\frac{\prod_{i=1}^s(z^{p_i}-1)}{(z-1)^{s-2}},
\end{equation}
where $s$ is the number of branches of the associated star graph
and $p_i$ are the number of nodes in the $i$--th branch (counting the vertex node), while $p\equiv\mathrm{l.c.m.}\{p_i\}$.

\paragraph{Auto--equivalences of the derived category.} Suppose \begin{equation}K\colon \mathsf{D}^b\,\mathsf{coh}\,\ce/\Z_p\to \mathsf{D}^b\,\mathsf{coh}\,\ce/\Z_p\end{equation} is an auto--equivalence (of triangulated categories).
$K$ induces an automorphism $\boldsymbol{K}$ of the Grothendieck group $\Gamma$
\begin{equation}\boldsymbol{K}\colon\Gamma\to \Gamma, \quad\text{given by}\quad [KX]=\boldsymbol{K}\cdot[X].\end{equation}
Choosing a $\Z$--basis in $\Gamma$,
$\bphi$ and $\boldsymbol{K}$ may be seen as $\mathrm{rank}\,\Gamma\times \mathrm{rank}\,\Gamma$ matrices with integral entries. For all auto--equivalences $K$ we have  $\tau K=K\tau$ (since the Auslander--Reiten translation is unique). Then $\bphi$ and $\boldsymbol{K}$, as matrices, commute
\begin{equation}\boldsymbol{K}\bphi=\bphi\boldsymbol{K},\end{equation}
and, in particular, $\boldsymbol{K}$ preserves the radical sublattice $\mathrm{rad}\,q\subset \Gamma$; this already follows from the fact that all auto--equivalences are isometries of the Euler form
\begin{equation}\langle \boldsymbol{K}\cdot[X],\boldsymbol{K}\cdot[Y]\rangle_E=\langle [X],[Y]\rangle_E.\end{equation}
 Comparing with \eqref{mattercharges}, we see that all auto--equivalence $K$ of the derived category $\mathsf{D}^b\,\mathsf{coh}\,\ce/\Z_p$ induces a reduced additive map
 \begin{equation}\overline{\boldsymbol{K}}\colon\Gamma_\mathrm{matter}\to \Gamma_\mathrm{matter}\end{equation}
 which commutes with the reduced Coxeter element $\overline{\bphi}$ and is an isometry of the reduced Tits form $\overline{q}(\cdot)$
 \begin{equation}\overline{q}\big(\overline{\boldsymbol{K}}\cdot f\big)=\overline{q}\big(f\big),\qquad\forall\; f\in\Gamma_\mathrm{matter}.\end{equation}

It follows that the image of the automorphism group of the derived category, $$\mathrm{Aut}\big(\mathsf{D}^b\,\mathsf{coh}\,\ce/\Z_p\big),$$ under the
homomorphism $\varrho\colon K\mapsto \overline{\boldsymbol{K}}$ is a subgroup of the \emph{finite} group $O(\Gamma_\mathfrak{g})$ of the $\Z$--isometries of the positive--definite Tits form $q_\mathfrak{g}$. More precisely, the image is a subgroup of the centralizer of the reduced Coxeter element
$\overline{\bphi}$
\begin{equation}\label{centralizer}\varrho\big(\mathrm{Aut}(\mathsf{D}^b\,\mathsf{coh}\,\ce/\Z_p)\big) \subset Z(\overline{\bphi})\subset O(\Gamma_\mathfrak{g})= \mathrm{Weyl}(\mathfrak{g})\ltimes \mathrm{Aut}(D_\mathfrak{g}),
\end{equation}
where $D_\mathfrak{g}$ is the  Dynkin graph of $\mathfrak{g}$.

\subparagraph{Remark.} The triality  of $\mathfrak{so}(8)$ is a group
of outer automorphisms; from the point of view of eqn.\eqref{centralizer} this means that its image is not contained in $\mathrm{Weyl}(\mathfrak{so}(8))$. For, say, $\mathfrak{g}=E_8$, $\mathrm{Aut}(D_\mathfrak{g})$ is trivial and the image of $\mathrm{Aut}(\mathsf{D}^b\,\mathsf{coh}\,\ce/\Z_p)$ is a subgroup of the Weyl group. The reader then may wonder in which sense the $S$--duality action is a generalization of triality which is an \emph{outer} action. The point is that the duality action is outer with respect to the natural group of `inner' automorphisms which is $\mathscr{W}$ (see eqn.\eqref{WW}). For SQCD $N_f=4$
$\mathscr{W}$ is the full Weyl group but it is a small subgroup for $E_6,E_7,E_8$.

\subsection{Coherent sheaves on weighted projective lines}\label{wpl}
With the exclusion\footnote{ The case of $\cn=2^*$, i.e.\! $p=1$, is rather similar. Indeed, the theory of coherent sheaves on the weighted projective lines of tubular type was constructed by Geigle and Lenzing in \cite{GL1} using as a model the Atiyah description of $\mathsf{coh}\,\ce$  \cite{aty}. The main technical difference is that for $p=1$ the canonical sheaf is trivial, while for $p>1$ is a $p$--torsion sheaf. Then for $p=1$ there is no tilting object.} of $\cn=2^*$, the complete $\cn=2$ gauge theories with gauge group $SU(2)$ are in one--to--one correspondence with the weighted projective lines having non--negative Euler characteristic $\chi\geq 0$ \cite{CV11,cattoy}. 
$SU(2)$ SYM coupled to a set of AD matter systems of types $D_{p_i}$ ($i=1,..,s$) corresponds to the weighted projective line with weights $(\boldsymbol{p})=(p_1,p_2,\dots, p_s)$.
The superconformal theories in this class are precisely the ones associated with the $\chi=0$ weighted projective lines\footnote{ Indeed, the coefficient of the $\beta$--function of the Yang--Mills coupling, $g_\mathrm{YM}$, is $-2\chi$, see \cite{CV11,cattoy,CDZG}.}; there are  four such lines with weights
\begin{equation}\label{tubular}(\boldsymbol{p})=(2,2,2,2),\quad(3,3,3),\quad(2,4,4),\quad
(2,3,6),
\end{equation}
the $i$--th weight $p_i$ being equal to the number of nodes in the $i$--th branch of the corresponding affine star graph (counting the vertex). The $\chi=0$ weighted projective lines
will be written
$\mathbb{X}_p$ (where $p\equiv\mathrm{l.c.m.}(p_i)=2,3,4,6$) or simply $\mathbb{X}$. We have \cite{GL1,handFM,equivalenceorbifold}
\begin{equation}\mathbb{X}_p= \ce/\Z_p,\end{equation}
and passing from elliptic orbifolds $\ce/\Z_p$ to weighted projective lines $\mathbb{X}_p$ is just a convenient shift in language.

\paragraph{Weighted projective lines {\rm \cite{GL1,lenzing1,lenzing2,lenzingmeltzer,meltzer,dirk,revLINE}}.} Given a set of positive integral weights\footnote{ For definiteness we write the $p_i$'s in a non--decreasing order.
Without loss we may assume $p_i\geq 2$.} $\boldsymbol{p}=(p_1,p_2,\dots,p_s)$ we define $L(\boldsymbol{p})$ to be the Abelian group
 over the generators
$\vec x_1,\vec x_2,\dots, \vec x_s$
subjected to the relations
\begin{equation}\vec c= p_1\vec x_1=p_2\vec x_2=\cdots=
p_s\vec x_s.\end{equation}
$\vec c$ is called the \emph{canonical} element of 
$L(\boldsymbol{p})$, while the \emph{dual} element is 
\begin{equation}
\vec \omega=(s-2)\vec c-\sum_{i=1}^s \vec x_i\in L(\boldsymbol{p}).\end{equation}
Given the weights $\boldsymbol{p}$ and $s$
distinct points $(\lambda_i:\mu_i)\in\mathbb{P}^1$ we define a ring
graded by $L(\boldsymbol{p})$
\begin{equation}S(\boldsymbol{p})=\bigoplus_{\vec a\in L(\boldsymbol{p})} S_{\vec a}= \C[X_1,X_2,\cdots, X_s,u,v]\Big/\big(X_1^{p_1}-\lambda_1 u -\mu_1 v, \;\cdots,\, X_s^{p_s}-\lambda_s u -\mu_s v\big)\end{equation}
where the degree of $X_i$ is $\vec x_i$
and the degree of $u,v$ is $\vec c$.
The weighted projective line $\mathbb{X}(\boldsymbol{p})$ is defined to be the projective scheme $\mathsf{Proj}\,S(\boldsymbol{p})$. Its Euler characteristic is \begin{equation}\chi(\boldsymbol{p})=2-\sum_{i=1}^s(1-1/p_i).\end{equation}
The Picard group of $\mathbb{X}(\boldsymbol{p})$ (i.e.\! the group of its invertible coherent sheaves $\equiv$ line bundles) is isomorphic to the 
group $L(\boldsymbol{p})$
\begin{equation}\mathsf{Pic}\,\mathbb{X}(\boldsymbol{p})=
\big\{\co(\vec a)\;\big|\; \vec a\in L(\boldsymbol{p})\big\},\end{equation}
i.e.\! all line bundles are obtained from the structure sheaf $\co\equiv \co(0)$ by shifting its degree in $L(\boldsymbol{p})$.
The dualizing sheaf is $\co(\vec\omega)$.
Hence
\begin{equation}\tau\, \co(\vec a)=\co(\vec a+\vec\omega).\end{equation}
One has
\begin{equation}\mathrm{Hom}(\co(\vec a),\co(\vec b))\simeq S_{\vec b-\vec a},\qquad \mathrm{Ext}^1(\co(\vec a),\co(\vec b))\simeq D\,S_{\vec a+\vec\omega-\vec b}.
\end{equation}
Any non--zero morphism between 
line bundles is a monomorphism \cite{lenzing1,lenzing2}. In particular, for all line bundles $L$,
 $\mathrm{End}\,L=\C$. Hence, if
$(\lambda:\mu)\in\mathbb{P}^1$ is \underline{not} one of the special $s$ points $(\lambda_i:\mu_i)$,
we have the exact sequence
\begin{equation}0\to \co\xrightarrow{\lambda u+\mu v}\co(\vec c)\to \cs_{(\lambda:\mu)}\to 0\end{equation}
which defines a coherent sheaf $\cs_{(\lambda:\mu)}$ concentrated at $(\lambda:\mu)\in\mathbb{P}^1$. It is a simple object in the category $\mathsf{coh}\,\mathbb{X}(\boldsymbol{p})$
(the `skyscraper').
At the special points $(\lambda_i:\mu_i)\in\mathbb{P}^1$
the skyscraper is not a simple object but rather it is an indecomposable of length $p_i$. The simple sheaves localized at the $i$--th special point $(\lambda_i:\mu_i)$
are the $\cs_{i,j}$ (where $j\in\Z/p_i\Z$)
defined by the exact sequences
\begin{equation}0\to \co(j\vec x_i)\to \co((j+1)\vec x_i)\to \cs_{i,j}\to 0.\end{equation}
Applying $\tau$ to these sequences we get 
\begin{equation}\tau \cs_{(\lambda;\mu)}=\cs_{(\lambda;\mu)},\qquad \tau\cs_{i,j}=\cs_{i,j-1}.\end{equation}

In conclusion we have\footnote{ The notation in the \textsc{rhs} \cite{lenzing1,lenzing2,ringel} stands for two properties: \textit{(i)} all object $X$ of 
$\mathsf{coh}\,\mathbb{X}(\boldsymbol{p})$ has the form $X_+\oplus X_0$ with $X_+\in\ch_+$, $X_0\in\ch_0$, and \textit{(ii)}
$\mathrm{Hom}(\ch_0,\ch_+)=0$.} \cite{lenzing1,lenzing2}
\begin{equation}\mathsf{coh}\,\mathbb{X}(\boldsymbol{p})=\ch_+\vee \ch_0,
\end{equation}
 where $\ch_0$ is the full Abelian subcategory of finite length objects (which is a
 uniserial category) and $\ch_+$
 is the subcategory of \emph{bundles}. 
 Any non--zero morphism from a line bundle $L$ to a bundle $E$ is a monomorphism. For all bundles $E$ we have a filtration\cite{lenzing1,lenzing2}
 \begin{equation}\label{filtration}0=E_0\subset E_1\subset E_2\subset\cdots\subset E_\ell=E,\end{equation}
 with $E_{i+1}/E_i$ line bundles. Then
 we have an additive function $\mathsf{rank}\colon K_0(\mathsf{coh}\,\mathbb{X}(\boldsymbol{p}))\to \Z$, the \emph{rank}, which is $\tau$--invariant, zero on $\ch_0$
 and positive on $\ch_+$. $\mathsf{rank}\,E$ is the length $\ell$ of the filtration \eqref{filtration}; line bundles have
 rank 1. 
 
 In physical terms \cite{cattoy} $\ch_0$ is the `light category' which encodes the zero   Yang--Mills coupling limit $g_\mathrm{YM}\to 0$; hence $\ch_0$ is well understood in terms of `perturbative' physics \cite{cattoy}.
 
 We define the additive function \emph{degree}, $\mathsf{deg}\colon K_0(\mathsf{coh}\,\mathbb{X}(\boldsymbol{p}))\to \frac{1}{p}\Z$, by
 \begin{equation}\mathsf{deg}\,\co\!\left(\sum\nolimits_i n_i\vec x_i\right)=\sum_i\frac{n_i}{p_i}.\end{equation}
$\mathsf{deg}$ satisfies the four properties:
 \textit{(i)} the degree is $\tau$ stable;
\textit{(ii)}  $\mathsf{deg}\,\co=0$; \textit{(iii)}  if $\cs$ is a simple of $\tau$--period $q$ one has $\mathsf{deg}\,\cs=1/q$; \textit{(iv)} $\mathsf{deg}\,X>0$ for all non--zero objects in $\ch_0$.

Physically, $\mathsf{rank}$ is the Yang--Mills
magnetic charge while $\mathsf{deg}$
is (a linear combination of) the Yang--Mills electric charge (and matter charges)
normalized so that the $W$ boson has charge $+1$. For the four weighted projective lines $\mathbb{X}_p$ with $\chi(\boldsymbol{p})=0$, eqn.\eqref{tubular}, the Riemann--Roch theorem reduces to the equality \cite{GL1,lenzing1,lenzing2}
\begin{equation}\label{RR}\frac{1}{p}\sum_{j=0}^{p-1}\big\langle [\tau^jX],[Y]\big\rangle_E =\mathsf{rank}\,X\,\mathsf{deg}\,Y-\mathsf{deg}\,X\,\mathsf{rank}\,Y.\end{equation}

\paragraph{Explicit formulae in the canonical basis.}
To write explicit expressions, it is convenient to choose a set of homological generators of $\mathsf{coh}\,\mathbb{X}(\boldsymbol{p})$; their classes then give
a $\Z$--basis of the Grothendieck group
$K_0(\mathsf{coh}\,\mathbb{X}(\boldsymbol{p}))$. It is convenient to choose the generators to be the direct summands of a tilting object of $\mathsf{coh}\,\mathbb{X}(\boldsymbol{p})$ \cite{GL1,lenzing1,lenzing2}. We choose
the \emph{canonical}\footnote{ \label{giustification}In view of \cite{cattoy} the existence of this tilting object justifies our claim that the BPS particles of the relevant QFT correspond to stable objects of the derived category 
$\mathsf{D}^b\mathsf{coh}\,\mathbb{X}(\boldsymbol{p})\equiv
\mathsf{D}^b\mathsf{mod}\,\Lambda(\boldsymbol{p})$. A more detailed analysis is presented in appendix \ref{tilting}. } such tilting object whose endomorphism algebra is the Ringel canonical algebra $\Lambda(\boldsymbol{p})$ of type $(\boldsymbol{p})$  \cite{GL1,lenzing1,lenzing2}. The canonical generating set consists of the following $n\equiv \sum_i (p_i-1)+2$ line bundles
\begin{equation}\label{bases}\co,\qquad \co(\ell\vec x_i)\ (\text{with }i=1,\dots,s,\ \ell=1,\dots, p_i-1),\qquad \co(\vec c).\end{equation}
By definition of tilting object, $\mathrm{Ext}^1$ vanishes between any pair of sheaves in eqn.\eqref{bases}, while
the only non--zero Hom spaces are
\begin{equation}\dim \mathrm{Hom}(\co,\co(\vec c))=2,\qquad
\dim \mathrm{Hom}(\co(k_i\vec x_i),\co(\ell_i\vec x_i))=1,\quad 0\leq k_i\leq \ell_i\leq p_i,\end{equation}
where, for all $i$, $\co(0\,\vec x_i)\equiv \co$ and $\co(p_i\vec x_i)\equiv \co(\vec c)$.

We write $(\phi_1,\dots, \phi_n)$ for the elements of the basis of $K_0(\mathsf{coh}\,\mathbb{X}(\boldsymbol{p}))$ 
 given by the Grothendieck classes of the $n$ line bundles in eqn.\eqref{bases} ordered so that the $\phi_1=[\co]$, $\phi_n=[\co(\vec c)]$ while
the $\{\phi_a\}_{a=2}^{n-1}$ are the $[\co(j\vec x_i)]$ listed in the $(i,j)$ lexicographic order.
The YM magnetic and electric charges of the generating sheaves \eqref{bases} are  
\begin{gather}\label{V1}(\mathsf{rank}\,\phi_1,\dots,\mathsf{rank}\,\phi_n)=(1,1,\dots,1,1)\equiv M^t\\
p(\mathsf{deg}\,\phi_1,\dots,\mathsf{deg}\,\phi_n)=(0,q_{1,1},q_{1,2},\dots,q_{1,p_1-1},\dots\dots, q_{s,1},q_{s,2},\dots,q_{s,p_s-1},p)\equiv Q^t.\label{V2}\end{gather}

\subparagraph{Specializing to the $\chi(\boldsymbol{p})=0$ case.} For the $\chi(\boldsymbol{p})=0$ weighted projective lines, eqn.\eqref{tubular},
the electric charges \begin{equation}
q_{i,1},\cdots, q_{i,p_i-1},p
\end{equation} of the sheaves \eqref{bases} are just
 the Coxeter labels on the $i$--th branch of the  associated affine star Dynkin graph $\mathfrak{g}^{(1)}$ numbered in increasing order from the most peripheral node to the vertex of the star (which has label $p$).  Explicitly,
\begin{equation}
q_{i,j}= \frac{p}{p_i}\,j.
\end{equation}
The Euler 
form between the elements
of the basis $E_{ab}=\langle\phi_a,\phi_b\rangle_E$  is given by the unipotent (upper triangular) block matrix
\begin{equation}\label{Eexplicit}E=\left(\begin{array}{c|c|c|c|c}
1 & 1_{p_1}^t & \cdots & 1_{p_s}^t & 2 \\\hline
0 & T_{p_1} & \cdots & 0 &1_{p_1}\\\hline
\vdots & \vdots & \ddots & \vdots &\vdots\\\hline
0 & 0 & \cdots & T_{p_s}  &1_{p_s}\\\hline
0 & 0 & \cdots &0 &1\end{array}
\right)\end{equation}
where $1_p$ stands for the column $(p-1)$--vector with 1's in all entries, and $T_p$ is the $(p-1)\times (p-1)$ triangular matrix with $1$'s along the main diagonal and everywhere above it.

In the canonical basis the Coxeter element is
represented by the matrix $\bphi_{ab}$ 
such that $\tau \phi_a=\bphi_{ab}\,\phi_b$; comparing with eqns.\eqref{A1}\eqref{A2} we get
\begin{equation}
\bphi=-E(E^t)^{-1}.
\end{equation} 

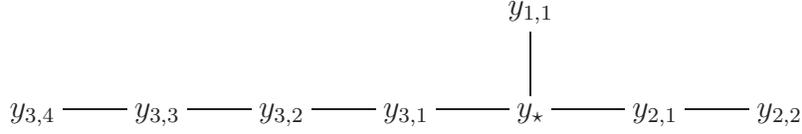
\begin{figure}
$$\xymatrix{&&&&y_{1,1}\ar@{-}[d] & \\
y_{3,4}\ar@{-}[r] &y_{3,3}\ar@{-}[r]& y_{3,2}\ar@{-}[r]&y_{3,1}\ar@{-}[r] & y_\star & y_{2,1}\ar@{-}[l] \ar@{-}[r]& y_{2,2}}$$
\caption{The assignments of integral variables $(\boldsymbol{y})$ to the nodes of the $E_8$ graph. For
$D_4,E_6,E_7$ just restrict to the corresponding Dynkin subgraph.\label{e8ex}}
\end{figure}

The radical of the Tits form $q$ is generated by the two vectors
\begin{align}\label{R1}R_1=&E^{-1}M\equiv 
(-1,0,\cdots,0,1)^t,\\
R_2=&E^{-1}Q.\label{R2}\end{align}
In particular, we note that the last two entries of the radical vector $E^{-1}(pM-Q)$ are $(\cdots, 1,0)^t$. It follows that in this basis we may identify the matter charge
lattice $\Gamma_\mathrm{matter}\equiv \Gamma/\mathrm{rad}\,q$ with the 
sublattice $\widehat{\Gamma}_\mathrm{matter}\subset\Gamma$ of vectors of the form
\begin{equation}\label{whichexvect}\boldsymbol{x}\equiv (x_\star,x_{1,1},\cdots,x_{1,p_1-1},\cdots\cdots, x_{s,1},\cdots, x_{s,p-2},0,0)\subset \Z^n\simeq \Gamma.\end{equation}
Since $E$ is upper triangular and unimodular, it maps a basis of the sublattice $\widehat{\Gamma}_\mathrm{matter}$ into a $\Z$--equivalent basis. Then we write the vectors \eqref{whichexvect} in the form
\begin{equation}
\boldsymbol{x}=E^{-1}\,\boldsymbol{y},\qquad \boldsymbol{y}\in\widehat{\Gamma}_\text{matter}.
\end{equation}
Let $\dot{E}$ (resp.\! $\ddot{E}$) be the principal submatrix of $E$ obtained by omitting the last (resp.\! the last two) row(s) and column(s). Essentially by definition,
\begin{align}
\dot{E}^{-1}+(\dot{E}^{-1})^t=C_{\mathfrak{g}^{(1)}}&=\left[{\text{the Cartan matrix of the}\atop \text{affine Lie algebra }\mathfrak{g}^{(1)}\phantom{nn}}\right.
\\
\ddot{E}^{-1}+(\ddot{E}^{-1})^t=C_{\mathfrak{g}}&=\left[{\text{the Cartan matrix of the}\atop \text{finite--type Lie algebra }\mathfrak{g}.}\right.
\end{align}
Then the restricted Tits form $\overline{q}$
on $\Gamma_\mathrm{matter}\simeq\widehat{\Gamma}_\mathrm{matter}$ is
\begin{equation}\overline{q}(\boldsymbol{x})=q_\mathfrak{g}\!\big(\boldsymbol{y}\big),\end{equation}
where $q_\mathfrak{g}(\boldsymbol{y})=
\frac{1}{2}\boldsymbol{y}^tC_\mathfrak{g}\boldsymbol{y}$ is the Tits form of the finite--dimensional Lie algebra $\mathfrak{g}$. The integers $(\boldsymbol{y})=(y_\star, y_{i,j})$  are attached to the vertices of the Dynkin graph $\mathfrak{g}$
as in figure \ref{e8ex}. In particular, $\overline{q}$ and $q_\mathfrak{g}$ are $\Z$--equivalent; the isometry between the two Tits forms is just multiplication by $E^{-1}$.
Under this isometry
\begin{equation}\widehat{\Gamma}_\mathrm{matter}\cong \Gamma_\mathfrak{g}.\end{equation}

\subparagraph{Remark.} Here we defined the matter charges as equivalence classes in $\Gamma/\mathrm{rad}\,q$. Of course, the physical matter charges are specific representatives of these classes. In \S.\ref{naive} we shall use the physical definition. However, the action of the $S$--duality group is independent of the choice of representatives, and often a different choice  simplifies the computations.

\paragraph{Structure of the derived category $\mathsf{D}^b\,\mathsf{coh}\,\mathbb{X}(\boldsymbol{p})$.}
The Abelian category $\ch=\mathsf{coh}\,\mathbb{X}(\boldsymbol{p})$ is \emph{hereditary}. The derived category
$\mathsf{D}^b\ch$ of a hereditary Abelian category $\ch$ coincides with its repetitive category $\mathsf{rep}\,\ch$ \cite{lenzing1,lenzing2}.
\medskip 

\noindent\textbf{Definition.} Let $\ca$ be an Abelian category.
 Its \emph{repetitive category} $\mathsf{rep}\,\ca$
 is
\begin{equation}\label{rep}\mathsf{rep}\,\ca= \bigvee_{n\in\Z} \ca[n],\end{equation}
whose objects are of the form 
\begin{equation}\label{sum}A=\bigoplus_{n\in\Z} A_n[n]\end{equation} with $A_n\in\ca$ and only finitely many $A_n$'s non--zero; the morphisms are
\begin{equation}
\mathrm{Hom}(A[m],B[n])=\mathrm{Ext}^{n-m}(A,B)
\end{equation}
   with Yoneda compositions. The notation in eqn.\eqref{rep} stands for eqn.\eqref{sum} together with the fact that $\mathrm{Hom}(A[m],B[n])=0$ for $m>n$. The
 translation functor is
   $A[m]\mapsto A[m+1]$.\medskip
   
 Thus, to describe $\mathsf{D}^b\ch$, it is enough to study the Abelian category $\ch$.  
 \medskip
 
 The \emph{slope} $\mu(E)$ of a coherent sheaf $E$ is the ratio of its degree and rank\footnote{ By convention, the zero object has all slopes.} 
 \begin{equation}\mu(E)=\mathsf{deg}\,E/\mathsf{rank}\,E.\end{equation} A coherent sheaf $E$ is \emph{semi--stable} (resp.\! \emph{stable}\footnote{ This notion of stability is related but \emph{distinct} from the one relevant for the existence of BPS states we mentioned at the beginning of this section. That notion is based on a stability function (the $\cn=2$ central charge $Z$) which depends on the couplings, masses and Coulomb branch point.}) if for all non--zero subsheaf $F$ one has
 $\mu(F)\leq \mu(E)$ (respectively $\mu(F)< \mu(E)$).
 
\subparagraph{The $\chi(\boldsymbol{p})=0$ case.} 
We restrict ourselves to the four weighted projective lines with $\chi=0$, i.e.\! to $\mathbb{X}_p$ with $p=2,3,4,6$. In this case all indecomposable coherent sheaf is semi--stable \cite{GL1,lenzing1}. 
 Let $\ch^{(q)}$ be the full (hereditary) Abelian subcategory of semi--stable objects of slope $q\in \mathbb{Q}\cup\infty$. For all $q$ one has $\ch^{(q)}\simeq \ch_0$ the full subcategory of finite--length objects. Then
\begin{gather}\label{herher}
 \ch=\bigvee_{q\in\mathbb{Q}\cup\infty} \ch^{(q)}\intertext{and hence} \mathsf{D}^b\,\ch=\bigvee_{n\in\Z}\bigvee_{q\in\mathbb{Q}\cup\infty} \ch^{(q)}[n],\label{herher2}
 \end{gather}
 where the notation
 implies that $\mathrm{Hom}_\ch(\ch^{(q)},\ch^{(q^\prime)})=0$ for $q>q^\prime$.
 $\ch_0$ (and then $\ch^{(q)}$ for all $q$) is a $\mathbb{P}^1$--family of stable tubes all of which are homogenous but the ones over the three (or four) special points $(\lambda_i:\mu_i)$ which have periods $p_i> 1$.  
 
 In conclusion, all indecomposable object $X$ of the derived category
 $\mathsf{D}^b\,\mathsf{coh}\,\mathbb{X}_p$ belongs to a $\ch^{(q)}[n]$ for some $q\in\overline{\mathbb{Q}}$ and $n\in\Z$;
 all these subcategories $\ch^{(q)}[n]$
 are equivalent to the `perturbative' category $\ch_0$ and hence physically well understood \cite{cattoy}.

\subsection{Telescopic functors and $\cb_3$ braid group action on $\mathsf{D}^b\,\mathsf{coh}\,\mathbb{X}_p$}\label{telescopic}

If our $\cn=2$ theory has $S$--duality,
the duality should be, in particular, a property of its BPS sector. Hence
the duality should act by automorphisms of the relevant derived category or, more precisely, of its orbit category \eqref{orbitcat}. For the five $SU(2)$ SCFTs we expect the $S$--duality group $\mathbb{S}$ to contain a  $PSL(2,\Z)$ from the internal torus of its 6d construction.
Before entering in the technical details, let us
see why this fact is rather natural in view of the peculiar form of the derived category as described in eqn.\eqref{herher2}.
We may identify $\overline{\mathbb{Q}}\equiv \mathbb{Q}\cup \infty$ with the projective line over the field $\mathbb{Q}$. The group $PSL(2,\Z)$ naturally acts on $\mathbb{P}^1(\mathbb{Q})$. Then the structure in the \textsc{rhs} of \eqref{herher2}
suggests the existence of an action of $PSL(2,\Z)$
on  $\mathsf{D}^b\,\mathsf{coh}\,\mathbb{X}_p$ which sends an object $X$ of slope $q\in\mathbb{P}^1(\mathbb{Q})$ into an object of slope $q^\prime=\gamma\cdot q\in\mathbb{P}^1(\mathbb{Q})$ for $\gamma\in PSL(2,\Z)$.
Since the slope is essentially the ratio of the Yang--Mills electric and magnetic charges,
such a natural $PSL(2,\Z)$ action will have the physical interpretation of the electro--magnetic $S$--duality. 

In order to implement this idea, we first need to normalize correctly the Yang--Mills charges so that they are integral, while the degree is quantized in units of $1/p$. So, for all object $X$ of the derived category, we set
\begin{equation}Y\!M(X)\equiv \begin{pmatrix}p\,\mathsf{deg}\,X\\
\mathsf{rank}\,X\end{pmatrix}\in\Z^2,\end{equation}
and call the integral $2$--vector $Y\!M(X)$ the \emph{Yang--Mills charges of} $X$.
Then we have to construct auto--equivalences of the derived category $\mathsf{D}^b\,\mathsf{coh}\,\mathbb{X}_p$ which induce on the Yang--Mills charges $Y\!M(\cdot)$ an action of $SL(2,\Z)$. 
To be concrete, consider the two matrices
\begin{equation}\label{2matrices}\mathsf{T}=\begin{pmatrix} 1 & 1\\0 & 1\end{pmatrix},\qquad 
\mathsf{L}=\begin{pmatrix} 1 & 0\\-1 & 1\end{pmatrix}.\end{equation}
$\mathsf{T}$ and $\mathsf{L}$ generate
$SL(2,\Z)$. Indeed, the braid group
on three strands, $\cb_3$, is generated by two elements $\mathsf{T}$, $\mathsf{L}$ subject to the single relation
\begin{equation}\label{braidrel}\mathsf{T}\mathsf{L}\mathsf{T}=\mathsf{L}\mathsf{T}\mathsf{L},\end{equation}
while we have
\begin{equation}\label{B2}
1\to Z(\cb_3)\to \cb_3\to PSL(2,\Z)\to 1,\end{equation}
where the center of the braid group,
$Z(\cb_3)$, is the infinite cyclic group generated by $(\mathsf{L}\mathsf{T})^3$. The two $2\times 2$ matrices in eqn.\eqref{2matrices}
satisfy the braid relation \eqref{braidrel} as well as $(\mathsf{L}\mathsf{T})^3=-1$, and hence generate the full $SL(2,Z)$.

To prove that the electro--magnetic duality group $SL(2,\Z)$ is part of
$\mathrm{Aut}(\mathsf{D}^b\,\mathsf{coh}\,\mathbb{X}_p)$ we need to construct two functors $T$ and $L$, which are auto--equivalences of the triangulated category $\mathsf{D}^b\,\mathsf{coh}\,\mathbb{X}_p$, and have the property that
\begin{equation}\label{whichproperty}Y\!M(T(X))=\mathsf{T}\,Y\!M(X),\qquad
Y\!M(L(X))=\mathsf{L}\,Y\!M(X),\end{equation}
so that the subgroup of $\mathrm{Aut}(\mathsf{D}^b\,\mathsf{coh}\,\mathbb{X}_p)$ generated by the two functors $T,L$ will induce an $SL(2,\Z)$ action on the Yang--Mills charges $Y\!M(\cdot)$.
Such auto--equivalences $T$, $L$ do exist: they are called
\emph{telescopic functors} \cite{lenzingmeltzer,meltzer,dirk}.

$T$ is simply the functor which shifts the $L(\boldsymbol{p})$ degree of the sheaf by $\vec x_3$\cite{lenzing2,meltzer,dirk}
\begin{equation}
X\longmapsto X(\vec x_3)\equiv T(X),\end{equation} where we ordered the weights so that $p_3\equiv p$ is the largest one. One can see the map
$X\longmapsto T(X)$ as the completion of a canonical map to a triangle of $\mathsf{D}^b\,\mathsf{coh}\,\mathbb{X}_p$ \cite{lenzingmeltzer,meltzer,dirk}
\begin{equation}\bigoplus_{j=0}^{p-1} \mathrm{Hom}^\bullet(\tau^j \cs_{3,0},X)\otimes \tau^j \cs_{3,0}\xrightarrow{\ \mathrm{can_X}\ } X\longrightarrow T(X),\end{equation}
so that $T$ induces the following action on the Grothendieck group $\Gamma$ \cite{dirk}
\begin{equation}[T(X)]=[X]-\sum_{j=0}^{p-1} \langle\cs_{3,j}, X\rangle_E\; [\cs_{3,j}].
\end{equation}
Explicitly, the action on the generating
set $\co$, $\cs_{i,j}$ is given by
\begin{equation}T(\co)=\co(\vec x_3),\quad T(\cs_{3,j})=\cs_{3,j+1}, \quad T(\cs_{i,j})=\cs_{i,j}\ \text{for }i\neq 3.\end{equation}
Thus $T$ preserves the $\mathsf{rank}$, while increases the \textsf{degree} by $1/p$ times the $\mathsf{rank}$; therefore
\begin{equation}
Y\!M(T(X))=\mathsf{T}\,Y\!M(X),\end{equation}
as required. The definition of the second functor $L$ is similar; one introduces the
triangle 
\begin{equation}\bigoplus_{j=0}^{p-1} \mathrm{Hom}^\bullet(\tau^j \co,X)\otimes \tau^j \co\xrightarrow{\ \mathrm{can_X}\ } X\longrightarrow L(X),\end{equation}
and proves that $X\longmapsto L(X)$ is an auto--equivalence of the derived category, see\! \cite{lenzingmeltzer,meltzer}.
The action of $L$ on the Grothendieck group is then
\begin{equation}\label{algo}[L(X)]=[X]-\sum_{j=0}^{p-1} \langle\tau^j\co, X\rangle_E\; [\tau^j\co].\end{equation}
This formula shows that $\mathsf{deg}\,L(X)=\mathsf{deg}\,X$ while
\begin{equation}\mathsf{rank}\,L(X)=\mathsf{rank}\,X-\sum_{j=0}^{p-1} \langle\tau^j\co,X\rangle_E=\mathsf{rank}\,X-p\,\mathsf{deg}\,X,
\end{equation}
where we used the Riemann--Roch theorem \eqref{RR}. Thus we get the desired property
\begin{equation}Y\!M(L(X))=\mathsf{L}\,Y\!M(X).\end{equation}

The explicit action on the generating
set $\co$, $\cs_{i,j}$ ($j=0,1,\dots,p_i-1$) is
\begin{align}L(\co)&=\tau^{-1}\co\equiv\co(-\vec \omega),\\
L(\cs_{i,j})&=\mathrm{ker}\!\!\left[\bigoplus_{k=1}^{p/p_i}\co\!\Big((k p_i-1-j)\vec \omega\Big)\xrightarrow{\ \text{can}\ }\cs_{i,j}\right]\!\![1],\end{align}
which, in the particular case $p_i=p$,
reduces to
\begin{equation}\label{speccase}L(\cs_{i,j})=\co\big(-\vec x_i+(p-1-j)\vec\omega\big)[1].\end{equation}

\paragraph{The  $\cb_3$ braid group relation.} It is easy to see that \cite{lenzingmeltzer,meltzer}

\begin{equation}\label{braidrel2}LTL=TLT.\end{equation}
As an illustration (and to establish a few useful equalities), we check that the two sides of the equality act in the same way on the structure sheaf $\co$ and on the simple sheaves $\cs_{3,j}$ (assuming $p_3=p$). By \textsc{proposition 5.3.4} of \cite{meltzer} this suffices to conclude that the two automorphisms of $\mathsf{D}^b\,\mathsf{coh}\,\mathbb{X}$ are at least \emph{isomorphic}. Since $\tau$ commutes with $T,L$, \begin{equation}LTL(\co)=LT(\tau^{-1}\co)=\tau^{-1}L(\co(\vec x_3)),\end{equation} while the triangle
\begin{equation}L(\cs_{3,0})[-1]\to L(\co)\to L(\co(\vec x_3))\end{equation}
gives 
\begin{equation}
L(\co(\vec x_3))=\cs_{3,0}\quad\Longrightarrow\quad LTL(\co)=\cs_{3,1}.\end{equation} On the other hand,
\begin{equation}\label{tlt1}TLT(\co)=TL(\co(\vec x_3))=T(\cs_{3,0})=\cs_{3,1}\equiv LTL(\co).\end{equation} By \eqref{speccase} we have 
\begin{equation}\label{tlt2}TLT(\cs_{3,j})=\tau^{-(j+2)}\co[1].\end{equation}
Eqn.\eqref{speccase} gives the triangle
\begin{equation}\begin{split}&\tau^{j+1}L(\cs_{3,j})[-1]\to \co\to \cs_{3,p-1}\\
&\xymatrix{\ar@{=>}[rr]^{\text{apply $\tau LT$}\atop \text{and rotate}} &&}
\quad
\tau L(\cs_{3,0})[-1]\to \tau^{j+2}\,LTL(\cs_{3,j})[-1]\to\cs_{3,p-1}.\end{split}\end{equation}
Since $\tau L(\cs_{3,0})[-1]\equiv\co(-\vec x_3)$ (cfr.\! \eqref{speccase}), the above triangle yields 
\begin{equation}\label{tlt3}LTL(\cs_{3,j})=\tau^{-(j+2)}\co[1]\end{equation}
 in agreement with eqns.\eqref{braidrel2}\eqref{tlt2}.
\medskip

Eqn.\eqref{braidrel2} says that the two functors $L$, $T$
generate a subgroup of $\mathrm{Aut}(\mathsf{D}^b\,\mathsf{coh}\,\mathbb{X})$
which is isomorphic to the braid group $\cb_3$. Note that the functor $TLT$ acts
on $Y\!M(X)$ by the matrix $\mathsf{S}$
\begin{equation}\mathsf{S}:=\mathsf{TLT}=\begin{pmatrix}0 & 1\\
-1 & 0\end{pmatrix}.\end{equation}
Arguing as in ref.\!\!\cite{cattoy} one concludes that
the functor $TLT$ has the physical effect of interchanging weak and strong Yang--Mills coupling $g_\mathrm{YM}\longleftrightarrow1/g_\mathrm{YM}$.

\paragraph{The center $Z(\cb_3)$ of the braid group.} 
In view of eqn.\eqref{B2}, we have $PSL(2,\Z)=\cb_3/Z(\cb_3)$. The center $Z(\cb_3)$ of $\cb_3$ is the infinite cyclic group generated by $(TL)^3$ which acts on the Yang--Mills charges as the non--trivial element of the center of $SL(2,\Z)$
 \begin{equation}(TL)^3\colon Y\!M(X)\longmapsto -Y\!M(X).\end{equation}
 
Using eqns.\eqref{tlt1}\eqref{tlt2}, we get
\begin{align}(TL)^3(\co)&=TLT\cdot LTL(\co)=
TLT(\cs_{3,1})=\tau^{-3}\co[1],\\
(TL)^3(\cs_{3,j})&=TLT\cdot LTL(\cs_{3,j})=\tau^{-(j+2)} TLT(\co)[1]=\tau^{-(j+2)}\cs_{3,1}[1]=\tau^{-3}\cs_{3,j}[1].\end{align}
So (again by \textsc{proposition 5.3.4} of\cite{meltzer}) we have the isomorphism of triangle functors
\begin{equation}(TL)^3\simeq \tau^{-3}\Sigma,\end{equation}
 where $\Sigma$ stands for the shift functor, $\Sigma(X)=X[1]$. Hence $(TL)^3$ induces an automorphism $(\bfT\bfL)^3$ of the Grothendieck lattice $\Gamma$ of order
 $2,2,4,2$ for $p=2,3,4,6$, respectively.
 
Consider the automorphism $\tau^3 (TL)^3\Sigma^{-1}$: \textit{(i)} it fixes the structure sheaf and the simples $\cs_{3,j}$,
\textit{(ii)}  it preserves the slope
$\mu(X)=\mathsf{deg}(X)/\mathsf{rank}(X)$,
and \textit{(iii)} its action on $\Gamma$ has order dividing
$2,2,4,2$, respectively.
The group of slope preserving automorphisms of $\mathsf{D}^b\,\mathsf{coh}\,\mathbb{X}$ which fix $\co$ is precisely $\mathrm{Aut}(\mathbb{X})$ i.e.\! (essentially) the permutations of equal weight special points in $\mathbb{P}^1$. Thus $\tau^3 (TL)^3\Sigma^{-1}$ is a permutation of the special points which respects their weights and leaves the third one fixed (by convention $p_3=p$). Hence $\tau^3 (TL)^3\Sigma^{-1}=1$ for $p=4,6$. For $p=2$, $\tau^3 (TL)^3\Sigma^{-1}$ is either the identity or an order 2 permutation of three objects $(x_1,x_2,x_4)$ which treats the three objects on the same footing; hence $\tau^3 (TL)^3\Sigma^{-1}=1$. For $p=3$
\begin{equation}\tau^3 (TL)^3\Sigma^{-1}\equiv (TL)^3\Sigma^{-1},\end{equation}
 is either the identity or the permutation $\pi_{12}$ of the first two special points.
 From their explicit action on the Grothendieck group, see \S.\ref{e6exp}, we conclude for the second
 possibility. Hence
 \begin{equation}\label{Wcenter}
 (TL)^3= \begin{cases}\tau^{-3}\Sigma &p\neq 3\\
 \pi_{12}\,\Sigma& p=3,
 \end{cases}
 \end{equation} 
where $\pi_{12}$ is the automorphism which interchanges the first two special points i.e.,\! in terms of the canonical generating set \eqref{bases},
$\co(\ell \vec x_1)\leftrightarrow \co(\ell\vec x_2)$ for $\ell=1,2$.

We note that $(TL)^3$ acts as $-1$ on the Yang--Mills charges and as $+1$ on the flavor charges.

\subsection{The groups $\mathrm{Aut}(\mathsf{D}^b\,\mathsf{coh}\,\mathbb{X})$ and $\mathbb{S}\equiv\mathrm{Aut}\,\cc(\mathbb{X})$} The subgroup $\cb_3$ generated by $L$, $T$ is not the full automorphism group
$\mathrm{Aut}(\mathsf{D}^b\mathsf{coh}\,\mathbb{X})$. We have a surjection of the automorphism group on $\cb_3$ whose kernel is the group of the automorphisms of the Abelian category $\mathsf{coh}\,\mathbb{X}$ which fix degree and rank. This is the group $\mathsf{Pic}(\mathbb{X})^0\ltimes \mathrm{Aut}(\mathbb{X})$, where 
$\mathsf{Pic}(\mathbb{X})^0$ is the group of degree zero line bundles and $\mathrm{Aut}(\mathbb{X})$ is the group of geometric automorphisms of $\mathbb{X}$, essentially\footnote{ For $D_4^{(1,1)}$ the situation is slightly subtler \cite{lenzingmeltzer,meltzer}. We shall ignore this aspect. } the group of permutations of the special points having the same weight $p_i$.
Thus, for a weighted projective line $\mathbb{X}$ having zero Euler characteristic,
$\chi(\mathbb{X})=0$, we have \cite{lenzingmeltzer,meltzer}
\begin{equation}\label{fullaut}1\to \mathsf{Pic}(\mathbb{X})^0\ltimes \mathrm{Aut}(\mathbb{X})\to 
\mathrm{Aut}(\mathsf{D}^b\mathsf{coh}\,\mathbb{X}) \to \cb_3\to 1.\end{equation}

In particular, the derived auto--equivalence
\begin{equation}\label{corre}
(TL)^3(\tau^{-1}\Sigma)^{-1}=
\begin{cases}\co(-2\vec\omega)\otimes - &p\neq 3\\
\pi_{12}\,\co(\vec\omega)\otimes - &p=3
\end{cases}\ \ \in\mathsf{Pic}(\mathbb{X})^0\ltimes \mathrm{Aut}(\mathbb{X}),
\end{equation}
belongs to the kernel of $\mathrm{Aut}(\mathsf{D}^b\mathsf{coh}\,\mathbb{X}) \to \cb_3$.

The cluster category of the weighted projective line $\mathbb{X}$ is defined to be the orbit category of $\mathsf{D}^b\,\mathsf{coh}\,\mathbb{X}$ with respect to the cyclic subgroup generated by $\tau^{-1}\Sigma$
\begin{equation}
\cc(\mathbb{X})=\mathsf{D}^b\,\mathsf{coh}\,\mathbb{X}\big/\big\langle \tau^{-1}\Sigma\big\rangle.
\end{equation}
Then comparing eqns.\eqref{fullaut},\eqref{B2}, and \eqref{corre} we get:
\medskip

 \textit{Let $\mathbb{X}$ be a weighted projective line with $\chi(\mathbb{X})=0$ and $\cc(\mathbb{X})=\mathsf{D}^b(\mathsf{coh}\,\mathbb{X}_2)\big/\langle \tau^{-1}\Sigma\rangle$ its cluster category. Then} (cfr.\! \textbf{Proposition 7.4} of \cite{BKL})
\begin{equation}1\to \mathsf{Pic}(\mathbb{X})^0\ltimes \mathrm{Aut}(\mathbb{X})\to \mathrm{Aut}(\cc(\mathbb{X}))\to PSL(2,\Z)\to 1.\end{equation}

We call the group
$\mathrm{Aut}\,\cc(\mathbb{X})$ the \emph{full $S$--duality group} written $\mathbb{S}$
\begin{equation}
\mathbb{S}=\big(\mathsf{Pic}(\mathbb{X})^0\ltimes \mathrm{Aut}(\mathbb{X})\big)\ltimes PSL(2,\Z).
\end{equation}
$\mathbb{S}$ is the $S$--duality group of the four SCFT $D_4^{(1,1)}$, $E_6^{(1,1)}$,
$E_7^{(1,1)}$ and $E^{(1,1)}_8$.
Its quotient group acting effectively on the Yang--Mills charges (modulo the Weyl action of the gauge group) is $PSL(2,\Z)$. The physical interpretation of the kernel subgroup 
\begin{equation}\label{WW}
\mathscr{W}\equiv\mathsf{Pic}(\mathbb{X})^0\ltimes \mathrm{Aut}(\mathbb{X})\end{equation} will be discussed in section 3.

\subsection{Some useful formulae}\label{useful}

Note that both our generating functors $L, T\in \mathrm{Aut}(\mathsf{D}^b\,\mathsf{coh}\,\mathbb{X})$ are of the form $X\longmapsto \cl_Y(X)$ where, for a fixed object $Y$,
the functor $\cl_Y$ is defined
by the triangle
\begin{equation}\bigoplus_{k=0}^{p-1}\mathrm{Hom}^\bullet(\tau^k Y,X)\otimes \tau^k Y\xrightarrow{\ \text{can}\ } X \longrightarrow \cl_Y(X).\end{equation}
Indeed one has
\begin{equation}
L=\cl_\co,\qquad T=\cl_{\cs_{3,0}}.
\end{equation}
A basic theorem (see \textsc{theorem 5.1.3} of \cite{meltzer}) states that, for a weighted projective line with $\chi(\mathbb{X})=0$, a functor $X\longmapsto \cl_Y(X)$
is an auto--equivalence of the derived category iff \textit{(i)} $p$ is length of the $\tau$--orbit of the object $Y$, and \textit{(ii)} the object $Y$ is \emph{quasi--simple}.  These two conditions imply, in particular,\footnote{ Here $\delta^{(p)}_{j,k}$ is the mod $p$ Kronecker delta, i.e.\! $\delta_{j,k}^{(p)}=1$ if $j=k\mod p$ and zero otherwise.} 
\begin{equation}\langle \tau^j Y,\tau^k Y\rangle_E=\dim \mathrm{Hom}(\tau^j Y,\tau^k Y)-\dim \mathrm{Hom}(\tau^k Y,\tau^{j+1} Y)=\delta_{j,k}^{(p)}-\delta_{j,k-1}^{(p)},\end{equation}
from which we get
\begin{equation}\label{norm1}\frac{1}{p}\sum_{j,k=0}^{p-1} \exp\!\big[2\pi i(j-k)s/p\big]\;\langle \tau^j Y,\tau^k Y\rangle_E=1-\exp\!\big[-2\pi i s/p\big].
\end{equation}
We shall see in \S.\ref{emergence} that this last equation guarantees that the induced action in the Grothendieck group
\begin{equation}\label{abstractaction}
[X]\longmapsto [\cl_Y,X]=[X]-\sum_{p=0}^{p-1}\big\langle [\tau^k Y],[X]\big\rangle_E\;[\tau^k Y],
\end{equation}
is an isometry of the Euler form i.e.
\begin{equation}
\big\langle \cl_Y X,\cl_Y Z\big\rangle_E=
\big\langle X,Z\big\rangle_E,
\end{equation}
which is (obviously) a necessary condition in order $\cl_Y$ to be an auto--equivalence.

\section{Warm--up: $SU(2)$ SQCD with $N_f=4$ again}

As a warm--up, we consider again $\cn=2$ SQCD with $G_\mathrm{gauge}=SU(2)$ and four flavors of quarks (the $D^{(1,1)}_4$ model). The eight quark states with
electric charge $+1$ transform in the vector representation of $SO(8)$. 
Seen as coherent sheaves on the weighted projective line $\mathbb{X}_2$ of weights $(2,2,2,2)$ these quark states correspond to the eight exceptional
simple sheaves
\begin{equation}\label{Quarks}\cs_{i,j}\qquad i=1,2,3,4,\ \ j=0,1.\end{equation} 
In the $\mathbb{X}_2$ case $\tau^2=\mathrm{Id}$, so $\bphi^2=1$. 
The Yang--Mills charges correspond to the $(+1)$--eigenvectors of
$\bphi$ and the flavor charges to the 
$(-1)$--eigenvectors. 

To make everything very transparent, in the next subsection we illustrate how flavor Spin(8) triality arise in concrete terms
i.e.\!\! writing down explicit expressions in the basis of $\Gamma_\mathrm{matter}$ which is standard in physics. Then in \S.\ref{f4root} we present a more elegant abstract viewpoint. The reader may prefer to skip
\S.\ref{naive}.

\subsection{Flavor charges and flavor weigths}\label{naive}

The four linear independent flavor charges are
\begin{align}\label{F1}
\alpha_1&=[\co(\vec x_3)]-[\co(\vec x_4)]=[\cs_{3,0}]-[\cs_{4,0}]\\
\alpha_2&=[\co(\vec x_2)]-[\co(\vec x_3)]=[\cs_{2,0}]-[\cs_{3,0}]\\
\alpha_3&=[\co]+[\co(\vec c)]-[\co(\vec x_1)]-[\co(\vec x_2)]=[\cs_{1,1}]-[\cs_{2,0}]\\
\alpha_4&=[\co(\vec x_1)]-[\co(\vec x_2)]=
[\cs_{1,0}]-[\cs_{2,0}]
\label{F2}\end{align}
i.e.\! the Grothendieck classes which have zero rank and zero degree.
Note that in the Grothendieck group we have the relation $[\cs_{i,0}]-[\cs_{i+1,0}]=[\cs_{i+1,1}]-[\cs_{i,1}]$ for all $i$. 
The Euler pairing restricted to the sublattice $\Gamma_\mathrm{flavor}\subset \Gamma$ generated by the charges $\{\alpha_a\}_{a=1}^4$ is 
\begin{equation}\label{dynkin}\langle \alpha_a,\alpha_b\rangle_E\equiv \alpha_a^t E\alpha_b= \text{the Cartan matrix of the $D_4$ graph}\quad
\begin{gathered}\xymatrix{& \alpha_1\ar@{-}[d]\\
\alpha_3\ar@{-}[r] & \alpha_2 \ar@{-}[r] &\alpha_4}\end{gathered}\end{equation}

$\Gamma_\mathrm{flavor}$ is thus identified with the root lattice of Spin(8).
An object $X$ of $\mathsf{D}^b\,\mathsf{coh}\,\mathbb{X}_2$ then carries a weight $(w_1(X),w_2(X),w_3(X),w_4(X))$ of the flavor Spin(8). Explicitly,
\begin{equation}\label{mainformula}
w_a(X)=\big\langle \alpha_a,[X]\big\rangle_E\equiv \big\langle [X],\alpha_a\big\rangle_E.
\end{equation} 
(Note that the Euler form is symmetric if one of its arguments is a flavor charge).

\begin{table}
\caption{Spin(8) weight of coherent sheaves on $\mathbb{X}_2$\label{tabtab}}
\begin{gather*}
\begin{tabular}{c|c||c|c}\hline\hline
sheaf & $(w_1,w_2,w_3,w_4)$ &
sheaf & $(w_1,w_2,w_3,w_4)$\\\hline\hline
$\cs_{1,0}$ & $(0,0,-1,1)$ & $\cs_{1,1}$ &
$(0,0,1,-1)$\\\hline
$\cs_{2,0}$ & $(0,1,-1,-1)$ & $\cs_{2,1}$
& $(0,-1,1,1)$\\\hline
$\cs_{3,0}$ & $(1,-1,0,0)$ & $\cs_{3,1}$
& $(-1,1,0,0)$\\\hline
$\cs_{4,0}$ & $(-1,0,0,0)$ & $\cs_{4,1}$
& $(1,0,0,0)$\\\hline\hline
\end{tabular}\\
\text{Spin(8) weights of exceptional simple sheaves}
\end{gather*} 
\begin{gather*}
\begin{tabular}{c|c||c|c}\hline\hline
sheaf & $(w_1,w_2,w_3,w_4)$ &
sheaf & $(w_1,w_2,w_3,w_4)$\\\hline\hline
$\co$ & $(0,0,1,0)$ & $\co(-\vec\omega)$ &
$(0,0,-1,0)$\\\hline
$\co(\vec x_1+\vec x_2-\vec c)$ & $(0,1,-1,0)$ & $\co(\vec x_3+\vec x_4-\vec c)$ & $(0,-1,1,0)$\\\hline
$\co(\vec x_1+\vec x_3-\vec c)$ & $(1,-1,0,1)$ & $\co(\vec x_2+\vec x_4-\vec c)$ & $(-1,1,0,-1)$\\\hline
$\co(\vec x_1+\vec x_4-\vec c)$ & $(-1,0,0,1)$ & $\co(\vec x_2+\vec x_3-\vec c)$ & $(1,0,0,-1)$\\\hline\hline
\end{tabular}\\
\text{Spin(8) weights of degree zero line bundles}
\end{gather*}
\begin{gather*}
\begin{tabular}{c|c||c|c}\hline\hline
sheaf & $(w_1,w_2,w_3,w_4)$ &
sheaf & $(w_1,w_2,w_3,w_4)$\\\hline\hline
$\co(\vec x_1)$ & $(0,0,0,1)$ & $\tau\co(\vec x_1)$ &
$(0,0,0,-1)$\\\hline
$\co(\vec x_2)$ & $(0,1,0,-1)$ & $\tau\co(\vec x_2)$
& $(0,-1,0,1)$\\\hline
$\co(\vec x_3)$ & $(1,-1,1,0)$ & $\tau\co(\vec x_3)$ & $(-1,1,-1,0)$\\\hline
$\co(\vec x_4)$ & $(-1,0,1,0)$ & $\tau\co(\vec x_4)$ & $(1,0,-1,0)$\\\hline\hline
\end{tabular}\\
\text{Spin(8) weights of degree 1 line bundles}
\end{gather*}
\end{table}

Using \eqref{mainformula} we easily compute the flavor weights of \textit{(i)} the exceptional simples \eqref{Quarks},
\textit{(ii)} the degree zero line bundles (which correspond to monopoles with magnetic charge $+1$ and zero electric charge), and
\textit{(iii)} the degree 1 line bundles (dyons with unit electric and magnetic charges), see
table \ref{tabtab}. We see that the 
quarks have the weights of the vector representation $\boldsymbol{v}$ of Spin(8),
the monopoles of the spinorial representation $\boldsymbol{s}$,
and the dyons of the spinorial representation $\boldsymbol{c}$. Of course, this is the physically correct result \cite{SW2}.

\paragraph{$PSL(2,\Z)$ and triality.}
Let us check how the two derived auto--equivalences $T$ and $L$ of \S.\ref{telescopic}
act on the Yang--Mills and flavor charges.
 $T$ acts on the generating set $\co$, $\cs_{i,j}$ as\footnote{ In this subsection we change our conventions from $T\colon X\to X(\vec x_3)$ to $T\colon X\to X(\vec x_1)$ to facilitate comparison with standard conventions in $SO(8)$ representation theory.}
\begin{equation}\label{onepoint}
\begin{aligned}T(\co)&=\co(\vec x_1),\\ T(\cs_{1,0})&=\cs_{1,1},\qquad
T(\cs_{1,1})=\cs_{1,0},\\
T(\cs_{i,j})&=\cs_{i,j}\hskip1.1cm i=2,3,4,\ \
j=0,1\end{aligned}\end{equation}
which is consistent with the expected action on the Yang--Mills charges\footnote{ Note that the factor 2 in the upper entry of the Yang--Mills charge vector for $N_f=4$ as compared to $\cn=2^*$ precisely corresponds to the discussion in \S.16 of \cite{SW2}. In the general case $2$ gets replaced by $p$.}
\begin{equation}\begin{pmatrix}2\,\mathsf{deg}\,T(X)\\
\mathsf{rank}\,T(X)
\end{pmatrix}=\begin{pmatrix} 1 & 1\\
0 & 1\end{pmatrix}\begin{pmatrix}2\,\mathsf{deg}\,X\\
\mathsf{rank}\,X
\end{pmatrix},\end{equation}
while the induced transformation on the Grothendieck group $\boldsymbol{T}\colon \Gamma\to\Gamma$ acts on the flavor charges $\alpha_a$ as (cfr.\! eqns.\eqref{F1}--\eqref{F2})
\begin{equation}\boldsymbol{T}(\alpha_1)=\alpha_1,\quad
\boldsymbol{T}(\alpha_2)=\alpha_2,\quad
\boldsymbol{T}(\alpha_3)=\alpha_4,\quad
\boldsymbol{T}(\alpha_4)=\alpha_3,
\end{equation}
in other words by the automorphism $\alpha_3\longleftrightarrow \alpha_4$ of the
Dynkin graph \eqref{dynkin}.
This is precisely the non--trivial element of the Spin(8)
triality group $\mathfrak{S}_3$ which maps the
weights of $\boldsymbol{v}$ into themselves, consistently with the fact that the $T$--duality makes sense at weak coupling where the quark fields keep their identity.

The functor $L$ acts as
\begin{equation}\label{whatL}
\begin{aligned}L(\co)&=\co(-\vec\omega),\\
L(\cs_{i,j})&=\co(-\vec x_i+(1-j)\vec\omega)[1],\qquad i=1,\dots,4,\ j=0,1.
\end{aligned}\end{equation}
Thus
\begin{equation}\begin{pmatrix}2\,\mathsf{deg}\,L(X)\\
\mathsf{rank}\,L(X)
\end{pmatrix}=\begin{pmatrix} 1 & 0\\
-1 & 1\end{pmatrix}\begin{pmatrix}2\,\mathsf{deg}\,X\\
\mathsf{rank}\,X
\end{pmatrix},\end{equation}
and (cfr.\! eqn.\eqref{algo})
\begin{equation}\label{LLL}
\boldsymbol{L}[\cs_{i,j}]=(-1)^{j-1}[\co(\vec x_i)]\mod\mathrm{rad}\,q,\qquad j=0,1,\end{equation}
so that the Spin(8) weights of the eight objects $L(\cs_{ij})$ are equal to those of the eight sheaves $\{\co(\vec x_i), \tau\co(\vec x_i)\}_{i=1}^4$ which are the weights of the spinorial representation $\boldsymbol{c}$.
Thus $L$ maps the Spin(8) vector representation
$\boldsymbol{v}$ into the spinorial representation $\boldsymbol{c}$.
However, the action of $L$ on the weights is not simply the automorphism $\alpha_1\longleftrightarrow \alpha_4$ of the Dynkin graph \eqref{dynkin} but rather the composition of this graph automorphism with a Spin(8) Weyl transformation.
In order to see this, we need to look at the full automorphism group
$\mathrm{Aut}(\mathsf{D}^b\,\mathsf{coh}\,\mathbb{X}_2)$.

From eqn.\eqref{fullaut} we know that, besides the braid group generated by
by $T$, $L$ we have the automorphism group $\mathsf{Pic}(\mathbb{X}_2)^0\ltimes \mathrm{Aut}(\mathbb{X}_2)$. Now
\begin{equation}\mathsf{Pic}(\mathbb{X}_2)^0=(\Z/2\Z)^4\big/(\Z/2\Z)_\mathrm{diag},\end{equation}
while, neglecting the subtlety already mentioned\footnote{ \label{subtle}The points is as follows: a permutation of the four special points does not give back the same $\mathbb{X}_2$ but rather a new one with
a different cross--ratio of the four special points. Since this cross--ratio does not enter anywhere in the BPS sector, as far as we are interested to BPS physics, we may identify all $\mathbb{X}_2$ and consider $\mathfrak{S}_4$ to be a symmetry of the physics. For a generic configuration of the four points the actual $\mathrm{Aut}(\mathbb{X}_2)$ is the Klein group. Hence the operation $\Pi$ defined in the text is an \emph{actual automorphism} of the derived category. }, we may effectively take $\mathrm{Aut}(\mathbb{X}_2)
\simeq \mathfrak{S}_4$, the permutation of the four special points in $\mathbb{P}^1$. Then
\begin{equation}\label{weyl8}\mathsf{Pic}(\mathbb{X}_2)^0\ltimes \mathrm{Aut}(\mathbb{X}_2)\simeq 
(\Z/2\Z)^4\big/(\Z/2\Z)_\mathrm{diag}
\ltimes \mathfrak{S}_4\simeq \mathrm{Weyl(Spin(8))}.\end{equation}

Let $\Pi$ be the autoequivalence of
the derived category associated with 
the permutation $\pi=(14)(23)$ of the four special points in $\mathbb{P}^1$. 
We define a new auto--equivalence of 
$\mathsf{D}^b\,\mathsf{coh}\,\mathbb{X}_2$
\begin{equation}\widehat{L}=\Pi L.\end{equation}
$\widehat{L}$ has the same action on the Yang--Mills charges as $L$. It follows from eqn.\eqref{LLL} that the Spin(8) weights of $\widehat{L}(\cs_{i,1})$ are equal to those of $\co(\vec x_{\pi(i)})$; comparing with table \ref{tabtab} we see that the effect of $\widehat{L}$ on the flavor weights is simply 
\begin{equation}\big(w_1(\widehat{L}(X)),w_2(\widehat{L}(X)), w_3(\widehat{L}(X)),w_4(\widehat{L}(X))\big)=\big(w_4(X),w_2(X), w_3(X),w_1(X)\big),
\end{equation}
i.e.\! $\widehat{L}$ induces the graph automorphism $\alpha_1\leftrightarrow \alpha_4$ which interchanges the Spin(8) spinorial representations $\boldsymbol{v}\leftrightarrow\boldsymbol{c}$.
However, $\widehat{L}$ is less convenient than $L$ since $\widehat{L}T\widehat{L}\neq T\widehat{L}T$.

\subsection{Relation with the $F_4$ root system}\label{f4root}

Let us look to the action of $\cb_3$ on the flavor charges from a more conceptual standpoint. We start from the formulae proven in \S.\ref{useful} which we
 specialize to $N_f=4$ SQCD. In this case both $\co$ and $\cs_{3,0}$ have $\tau$--period $p=2$. For $p=2$ the action of $\cl_Y$ on the Grothendieck group, eqn.\eqref{abstractaction}, reduces to 
\begin{equation}\begin{split}\cl_{Y} [X]&=[X]-\langle [Y],[X]\rangle_E\, [Y]-
\langle [\tau Y], [X]\rangle_E\, [\tau Y]=\\
&=[X]-\left\langle \frac{[Y]+[\tau Y]}{2}, [X]\right\rangle_E\;\big([Y]+[\tau Y]\big)
-\left\langle \frac{[Y]-[\tau Y]}{2}, [X]\right\rangle_E\;\big([Y]-[\tau Y]\big)
\end{split}\end{equation}
with $Y=\co$, $\cs_{3,0}$ ($\cs_{1,0}$) respectively. Since $[\tau Y]=\bphi\cdot[Y]$,
the charges $[Y]\pm [\tau Y]$ are $\pm 1$ eigenvectors of $\bphi$ i.e.\! they are a Yang Mills and a matter charge, respectively.
Hence, the induced action on the matter
lattice is $\Gamma_\mathrm{matter}\equiv \Gamma/\mathrm{rad}\,q$ is simply
\begin{equation}\overline{\cl}_{Y} [X]=[X]-\frac{1}{2}\Big\langle [Y]-[\tau Y], [X]\Big\rangle_{\!\mathrm{sym}}\;\big([Y]-[\tau Y]\big)
\end{equation}
where $\langle \cdot,\cdot \rangle_\text{sym}$ is the  inner product on $\Gamma_\text{matter}\simeq\Gamma_{D_4}$ given by the Cartan matrix.
The action of an automorphism is an isometry of the positive definite symmetric pairing; this requires
\begin{equation}\big\|[Y]-[\tau Y]\big\|^2\equiv \big\langle [Y]-[\tau Y],[Y]-[\tau Y]\big\rangle_E=4\end{equation}
 which is automatic in view of eqn.\eqref{norm1}.
From eqns.\eqref{fullaut}\eqref{weyl8} we conclude that
\medskip

\textit{$\mathrm{Aut}(\mathsf{D}^b\,\mathsf{coh}\,\mathbb{X}_2)$ acts\footnote{ Neglecting the subtlety in footnote \ref{subtle}.} on the flavor root lattice $\Gamma_{\mathfrak{so}(8)}$
as the group of reflections generated by
its elements of square--length $2$ and $4$. This group is isomorphic to $\mathrm{Weyl}(F_4)$ {\rm \cite{conway}}. One has
\begin{equation}1\to\mathrm{Weyl}(SO(8))\to \mathrm{Weyl}(F_4)\to \mathfrak{S}_3\to 1.\end{equation}}

The triality group is simply $\mathrm{Weyl}(F_4)/\mathrm{Weyl}(SO(8))$.

\subsection{The cluster category}

Around eqn.\eqref{orbitcat} we claimed that the triangulated category which describes the physics of 
the BPS sector is not $\mathsf{D}^b\,\mathsf{coh}\,\mathbb{X}$ itself but rather an orbit category of the form $\mathsf{D}^b\,\mathsf{coh}\,\mathbb{X}/\mathscr{C}$ where $\mathscr{C}$ is a certain subgroup of
 $\mathrm{Aut}(\mathsf{D}^b\,\mathsf{coh}\,\mathbb{X})$. The subgroup $\mathscr{C}$ is determined by non--perturbative physical considerations, in particular by the study of the quantum monodromy $\mathbb{M}(q)$ and its fractional powers \cite{CNV,Ysy}. In the case of a Lagrangian field theory, however,
the group $\mathscr{C}$ should also have an elementary
interpretation in terms of conventional weak--coupling physics. Then we use $N_f=4$ $SU(2)$ SQCD as a convenient example to shed light on the physical meaning of the cluster category.
 
 We claim that passing from $\mathsf{D}^b\,\mathsf{coh}\,\mathbb{X}_2$ to $\cc(\mathbb{X}_2)$
in this example corresponds to implementing the Gauss' law on the physical states.
Indeed, the center $Z(\cb_3)$ of the braid group $\cb_3$ is the infinite cyclic group generated by $(TL)^3=(LT)^3$. For $p=2$ one has the equality\begin{equation}(TL)^3\equiv \tau^{-3}\Sigma=\tau^{-1}\Sigma\end{equation} and hence $(TL)^3$ acts on the Yang--Mills charges $Y\!M(X)$ as multiplication by $-1$,
while fixing all flavor charges.
Thus $(TL)^3$ acts on $\Gamma$ as the
Weyl group of the gauge group $SU(2)$, which is a gauge transformation belonging to the connected component of the identity which leaves invariant all physical states satisfying the Gauss' law. Hence $(TL)^3\equiv \tau^{-1}\Sigma$ should act trivially on the physical states, i.e.\! two objects in the same $\tau^{-1}\Sigma$--orbit should be considered the same `physical' object. 
Then the natural triangulated category which is associated to the physical BPS states is
\begin{equation}\cc(\mathbb{X}_2)=\mathsf{D}^b(\mathsf{coh}\,\mathbb{X}_2)\big/\langle \tau^{-1}\Sigma\rangle,
\end{equation}
which is precisely the definition of the
cluster category for the coherent sheaves on the weighted projective line of type $(2,2,2,2)$ (equivalently of the canonical algebra of the same type) \cite{keller,BKL}.

It is convenient to see the orbit category $\cc_2\equiv\mathsf{D}^b(\mathsf{coh}\,\mathbb{X}_2)/\langle \tau^{-1}\Sigma\rangle$ as a (triangulated) category with the same objects as the triangulated category $\mathsf{D}^b(\mathsf{coh}\,\mathbb{X}_2)$ and morphism spaces 
\begin{equation}\label{eeeqref}
\mathrm{Hom}_{\cc_2}(X,Y)=\bigoplus_{g\in \langle \tau^{-1}\Sigma\rangle}
\mathrm{Hom}_{\mathsf{D}^b(\mathsf{coh}\,\mathbb{X}_2)}(X,g Y),
\end{equation}
see appendix \ref{tilting} for further details. Then the isoclasses of $\cc_2$ objects are the orbits of isoclasses of objects of the derived category, and the physical BPS states --- satisfying Gauss' law --- are in one--to--one correspondence with the  isoclasses of 
stable $\cc_2$ objects.
The physical meaning of equation \eqref{eeeqref} is that defining the ``physical'' morphism space to be the direct sum of morphisms in the original category from $X$ to $gY$ for all $g$, makes the morphism to account correctly for all the possible ``relative gauge orientations'' of the two objects when they form a BPS bound state.

\section{$S$--duality for the $E^{(1,1)}_r$ models}

\subsection{Generalities}
We have seen in \S.\ref{overview} that the matter charges of the five $\mathfrak{g}_r^{(1,1)}$ models ($\mathfrak{g}=A_1,D_4,E_6,E_7,E_8$) take values in the weight lattice
$\Gamma^\mathrm{w}_\mathfrak{g}$ of $\mathfrak{g}$. By this we mean that they are valued on a lattice endowed with the Cartan (integral) symmetric form $\langle\cdot,\cdot\rangle_\mathrm{sym}$ which is preserved by all 
symmetries and dualities of the quantum field theory.
On the root lattice $\Gamma_\mathfrak{g}$ there is, in addition, an integral skew--symmetric form
$\langle\cdot,\cdot\rangle_\text{Dirac}$ which is also preserved by all dualities. Indeed, 
the reduced Coxeter element $\overline{\bphi}$ centralizes the
action of all dualities on $\Gamma_\mathfrak{g}$, cfr.\! eqn.\eqref{centralizer}.
However there are two major differences between the Lagrangian models $A_1^{(1,1)}$, $D_4^{(1,1)}$ and the non--Lagrangian ones
$E_6^{(1,1)}$, $E_7^{(1,1)}$, $E_8^{(1,1)}$:
\begin{itemize}
\item[A)]
In the Lagrangian case the skew--symmetric form $\langle\cdot,\cdot\rangle_\text{Dirac}$ is identically zero on $\Gamma_\mathfrak{g}$, i.e.\! the matter charges are pure flavor ($\equiv$ the matter system are free hypermultiplets);
\item[B)] in the Lagrangian case the Lie group $\exp(\mathfrak{g})$ is a \emph{symmetry} of the theory. This is \underline{not} true in the non--Lagrangian case.
\end{itemize}
Indeed, in section 3 we saw that, in the two Lagrangian models, the class of coherent sheaves $X$ with a fixed Yang--Mills charge $Y\!M(X)\equiv (p\,\mathsf{deg}\,X,\mathsf{rank}\,X)^t$ have the correct $\mathfrak{g}$ weights to form a complete representation of $\mathfrak{g}$. This is certainly \underline{not} true for $p=3,4,6$. For instance, the number of zero--degree line bundles is equal to the order of the restricted Picard groups
\begin{equation}\mathsf{Pic}(\mathbb{X}_3)^0=\Z/3\Z\times \Z/3\Z,
\qquad\mathsf{Pic}(\mathbb{X}_4)^0=\Z/2\Z\times \Z/4\Z,
\qquad\mathsf{Pic}(\mathbb{X}_6)^0=\Z/2\Z \times \Z/3\Z,
\end{equation}
and no non--trivial representation of $E_6$, $E_7$ and $E_8$ have dimension $9$, $8$, and $6$. Alternatively we may note that $\mathrm{Weyl}(E_r)$ is \underline{not} a triangle auto--equivalence for $E_r^{(1,1)}$ models. The subgroups of $\mathrm{Aut}(\cc(\mathbb{X}_p))$ which fix the rank and the degree are listed in the second column of the following table:
\begin{equation}\begin{array}{l|l|l}
p=3 &(\Z/3\Z\times \Z/3\Z)\ltimes \mathfrak{S}_3 &SU(3)\\
p=4 &(\Z/2\Z\times \Z/4\Z)\ltimes \mathfrak{S}_2&Sp(2)\\
p=6 &\;\,\Z/2\Z\times \Z/3\Z& Sp(1)\end{array}
\end{equation}
They certainly do not contain the Weyl group of any Lie group but those listed in the third column and their subgroups.
The lattice of flavor charges
\begin{equation}\Gamma_\mathrm{fl}=\Big\{x\in\Gamma\;\big|\; \bphi\cdot x=-x\Big\},
\end{equation}
has rank $0,3,2$ for $p=3,4,6$. For $p=4$
$\Gamma_\mathrm{fl}$ is generated by the three classes
\begin{equation}\begin{aligned}
\alpha_1&=[\cs_{2,0}]+[\cs_{2,2}]-[\cs_{1,1}],\\\alpha_2&=[\cs_{3,0}]+[\cs_{3,2}]-[\cs_{1,0}],\\
\alpha_3&=[\cs_{1,0}]-[\cs_{2,0}]-[\cs_{2,2}]
\end{aligned}\end{equation}
The symmetric form on $\Gamma_\mathrm{fl}$
\begin{equation}
\langle\alpha_a,\alpha_b\rangle_E=\begin{pmatrix}3 & -1 & -1\\
-1 & 3 & -1\\ -1 & -1 & 3\end{pmatrix}
\end{equation}
is not related in any obvious way to a Cartan matrix, which is consistent with the fact that (contrary to the $p=2$ case studied in sect.\! 3) physically we do not expect any non--Abelian enhancement of the flavor symmetry for the $p=4$ model.
$\mathrm{Aut}(\bX_4)$ just permutes the two identical matter AD subsystems of type $D_4$,
and acts on the flavor charges as
\begin{equation}
\alpha_1\leftrightarrow \alpha_1+\alpha_2+\alpha_3,\qquad \alpha_2\leftrightarrow -\alpha_3.
\end{equation}
The two generators of $\mathsf{Pic}(\bX_4)^0$ act as\footnote{ A perhaps more transparent description of the action of the Picard group on the flavor charges is the following: $\co(\vec x_2-\vec x_3)$ inverst the signs of the flavor charges of the two $D_4$ matter systems, while $\co(\vec x_1-2\vec x_2)$ invert the sign of the flavor charge of the $D_2$ hypermultiplet.} 
\begin{align}
 \co(\vec x_2-\vec x_3)\colon (\alpha_1,\alpha_2,\alpha_3)&\mapsto (\alpha_3,-\alpha_1-\alpha_2-\alpha_3,\alpha_1)\\
\co(\vec x_1-\vec 2x_2)\colon (\alpha_1,\alpha_2,\alpha_3)&\mapsto (\alpha_3,\alpha_1+\alpha_2+\alpha_3,\alpha_1).
\end{align}
For $p=6$ the generators of $\Gamma_\mathrm{fl}$ are
\begin{equation}
\alpha_1=[\cs_{1,1}]-[\cs_{3,0}]-[\cs_{3,2}]-[\cs_{3,4}],\quad \alpha_2=-[\cs_{1,0}]+[\cs_{3,0}]+[\cs_{3,2}]+[\cs_{3,4}]
\end{equation}
with
\begin{equation}
\langle\alpha_a,\alpha_b\rangle_E=\begin{pmatrix} 4 &-2\\ -2 & 4\end{pmatrix}=2\, C_{A_2}.
\end{equation}
Tensoring with $\co(\vec x_1-3\vec x_3)$ changes the sign of both flavor charges, while
the flavor charges are inert under $\co(\vec x_2-2\vec x_3)$.

\subsection{Emergence of Shephard--Todd
complex reflection groups}\label{emergence}

From \S.\ref{useful} we know that the action of the two functors $L$ and $T$ on the Grothendieck group $\Gamma$ is given by
\begin{equation}[LX]=[X]-
\sum_{k=0}^{p-1}\big\langle\bphi^k\cdot [\co],[X]\big\rangle_E\; \bphi^k\cdot[\co]\end{equation}
\begin{equation}[TX]=[X]-
\sum_{k=0}^{p-1}\big\langle\bphi^k\cdot ([\co(\vec c)]-[\co((p-1)\vec x_3]),[X]\big\rangle_E\; \bphi^k\cdot([\co(\vec c)]-[\co((p-1)\vec x_3])\end{equation}
where we used 
\begin{equation}[\cs_{3,p-1}]=[\cs(\vec c)]-[\co((p-1)\vec x_3)].\end{equation}

Writing $[X]=\sum_a [X]_a\phi_a$, where
$\{\phi_a\}_{a=1}^n$ is the canonical basis of $\Gamma$ (cfr.\! \S.\ref{wpl}) and $[X]_a\in\Z$,
we have\footnote{ The matrices $\bfT$ and $\bfL$ are defined to act on the left of the basis elements $\phi_a\xrightarrow{\;L,T\;} \bfL_{ab}\,\phi_b,\; \bfT_{ab}\,\phi_b$, and hence they act on the \emph{right} of the coefficients $[X]_a$.}
\begin{align}\label{m1}[X]_b\,\boldsymbol{L}_{ba}&\equiv
[LX]_a=\delta_{ab}\, [X]_b-\sum_{k=0}^{p-1}(\bphi^k)_{1a}\,(\bphi^k)_{1c}E_{cb}[X]_b\\
[X]_b\,\boldsymbol{T}_{ba}&\equiv
[TX]_a=\delta_{ab}\, [X]_b-\sum_{k=0}^{p-1}\Big((\bphi^k)_{r\,a}-(\bphi^k)_{r-1\,a}\Big)\Big((\bphi^k)_{r\,c}-(\bphi^k)_{r-1\,c}\Big)E_{cb}[X]_b.\label{m2}
\end{align}
Since $\bphi=-E(E^{t})^{-1}$, the matrices
$\boldsymbol{T}$ and $\boldsymbol{L}$
are expressed in terms of the Euler matrix $E$ only; plugging in the explicit matrix $E$ in eqn.\eqref{Eexplicit}, we get  integral
$n\times n$ matrices $\bfT$ and $\bfL$ giving the concrete action of $T$, $L$ on the lattice $\Gamma$. See appendix \ref{explicitmatrices}.

For $p=3,4,6$, the matrices $\bfT$ and $\bfL$ give a (reducible) $n=8,9,10$ dimensional representation of $\cb_3$
on the vector space $V=\Gamma\otimes\C$.
Indeed they satisfy the braid relation
\begin{equation}\bfT\bfL\bfT=\bfL\bfT\bfL.\end{equation}
$\bfT$ and $\bfL$ commute with $\bphi$ and, for $p=3$, also with the permutation $\pi_{12}$ in eqn.\eqref{Wcenter}.
We decompose $V$ in eigenspaces of the semisimple linear map
$\bphi$,
\begin{equation}V=\bigoplus_{s=0}^{p-1} V_s,\qquad\quad 
\bphi\Big|_{V_s}=e^{2\pi i s/p}.\end{equation}
From eqns.\eqref{m1}\eqref{m2} the restrictions of the braid generators $\bfT$, $\bfL$ to each $\bphi$--eigenspace, 
\begin{equation}\bfL_s,\bfT_s\colon V_s\longrightarrow V_s\end{equation}
 are
\begin{align}\label{LT1}\bfL_s&=\boldsymbol{1}-\frac{1}{p}\left(\sum_{k=0}^{p-1} e^{-2\pi i k s/p}\,\bphi^kv\right) v^tE\left(\sum_{\ell=0}^{p-1} e^{-2\pi i \ell s/p}\,\bphi^\ell\right)\\
\bfT_s&=\boldsymbol{1}-\frac{1}{p}\left(\sum_{k=0}^{p-1} e^{-2\pi i k s/p}\,\bphi^kw\right) w^tE\left(\sum_{\ell=0}^{p-1} e^{-2\pi i \ell s/p}\,\bphi^\ell\right)\label{LT2}\\
&\text{where}\qquad v=(1,0,\cdots,0)^t,\qquad w=(0,\cdots, 0,-1,1)^t.\end{align}
Eqns.\eqref{LT1}\eqref{LT2} together with eqn.\eqref{norm1} imply that the \emph{minimal} equations satisfied by the restricted generators $\bfT_s,\bfL_s$ are\begin{equation}\label{resgen}
\begin{cases}(\bfL_s-1)(\bfL_s-e^{-2\pi is/p})=(\bfT_s-1)(\bfT_s-e^{-2\pi is/p})=0 &\text{if }\dim V_s>1\\
\bfL_s-e^{-2\pi is/p}=\bfT_s-e^{-2\pi is/p}=0
&\text{if }\dim V_s=1.
\end{cases}\end{equation}
In particular, acting on $V_0$ --- which corresponds to the Yang--Mills charges $Y\!M(\cdot)$ --- $\bfT_0$ and $\bfL_0$ are $2\times 2$ irreducible unipotent matrices in agreement with eqns.\eqref{whichproperty}\eqref{2matrices}. For $p$ even, $\bfT_{p/2}$ and $\bfL_{p/2}$ acting on $V_{p/2}$ --- i.e.\! on the flavor charges --- are involutions, in fact reflections on square--length $2p$ lattice vectors
\begin{equation}y\longmapsto y- \frac{1}{p} \langle u, y\rangle_\mathrm{sym}\, u,\quad \text{for some}\ \ u\in \Gamma,\ \|u\|^2=2p,\end{equation}
 just as found in \S.\ref{f4root} for the $D_4^{(1,1)}$ model.
 
 The minimal polynomial for the $n\times n$ matrices $\bfT$, $\bfL$ which give the action of the telescopic functors $T$, $L$ on the Grothendieck group $\Gamma$ is then
\begin{equation}
\big(\bfT-1\big)\big(\bfT^p-1)=\big(\bfL-1)\big(\bfL^p-1)=0.\label{mmma}
\end{equation}
Indeed, more generally, the following identities are true for \emph{all} analytic function
$f(z)$
\begin{equation}
\Big(f(\bfT)-f(\bphi^{-1})\Big)\Big(\bfT-1\Big)=\Big(f(\bfL)-f(\bphi^{-1})\Big)\Big(\bfL-1\Big)=0,
\end{equation}
eqn.\eqref{mmma} being the special case $f(z)=z^p$.
In addition, the following identities hold 
(see appendix \ref{explicitmatrices})
\begin{equation}
(\bfT^t)^p=1+R_1\otimes M^t\qquad
(\bfL^t)^p=1+R_2\otimes Q^t,
\end{equation}
where $M,Q,R_1,R_2$ are the magnetic/electric charge and radical vectors defined in eqns.\eqref{V1}, \eqref{V2}, \eqref{R1} and \eqref{R2}.

\paragraph{$\cb_3$ action on the matter charges.} The braid group action on the \emph{matter}  vector space
\begin{equation}V_\mathrm{matter}=\bigoplus_{s\not= 0}V_s\equiv \Gamma_\mathrm{matter}\otimes\C\end{equation} preserves the positive--definite Hermitian product induced by the Euler form as well as the lattice $\Gamma_\mathrm{matter}\subset V_\mathrm{matter}$.
Hence the quotient of $\cb_3$ which  acts effectively on $V_\mathrm{matter}$ is a \emph{finite} group $G_\text{matter}$. This group is generated by the restriction of the matrices $\bfT$ and $\bfL$ to the $(n-2)$--dimensional space $V_\text{matter}$. By abuse of notation, we denote these `matter' restrictions by the same symbols $\bfT$, $\bfL$.
Then the finite group $G_\text{matter}$ is generated by the two elements $\bfT$, $\bfL$
subjected
\underline{at least} to the obvious relations
\begin{equation}\label{atleast}
\bfL^p=\bfT^p=1,\quad \bfT\bfL\bfT=\bfL\bfT\bfL.\end{equation}
Each eigenspace $V_s$ with $s\not=0$ carries a unitary representation of $G_\text{matter}$ (with respect to the Hermitian product induced by the Euler form).
From eqns.\eqref{LT1}\eqref{LT2} we see that
\begin{equation}\label{refgr}\begin{gathered}\text{\textit{The multiplicity of $e^{-2\pi is/p}$ as an eigenvalue of $\bfT_s$ and $\bfL_s$ in $V_s$ is}}\\
\text{\textit{precisely one. Their other eigenvalues are all equal $1$ {\rm(cfr.\! eqn.\eqref{resgen}).}}}
\end{gathered}\end{equation}

Given the underling real structure on the
complex vector space $V_\text{matter}$ induced by the lattice $\Gamma_\text{matter}$ we know that the $G_\text{matter}$--modules
$V_s$ and $V_{p-s}$ are conjugate 
\begin{equation}V_{p-s}=\overline{V_s},\qquad p=3,4,6,\ \ \ s=1,2,\dots, p-1.\end{equation}

\paragraph{The case $V_s$ is one--dimensional.}
If $\dim V_s=1$ the action of both generators $\bfT_s$, $\bfL_s$ is just multiplication by $e^{-2\pi is/p}$, and the quotient group of $G_\mathrm{matter}$ which acts effectively on $V_s$ is the cyclic group
\begin{equation}\Z\Big/\tfrac{p}{\gcd(s,p)}\Z.\end{equation} 
In this case the action of the generator of $Z(\cb_3)$ on $V_s$ is 
\begin{equation}
(\bfT_s\bfL_s)^3= e^{-12\pi i s/p}\equiv \begin{cases}1 &\text{for }p=3,6\\
(-1)^s &\text{for }p=4.\end{cases}
\label{onedim1}
\end{equation}

From \S.\ref{telescopic} we know that
\begin{equation}
(\bfT\bfL)^3=\begin{cases}-\pi &p=3\\
-\bphi^{-3} & p=2,4,6.
\end{cases}
\end{equation}
Thus for $p\neq 3$ we have 
\begin{equation}\label{phiact}
(\bfT\bfL)^3\big|_{V_s}=-e^{-6\pi is/p},
\end{equation}
which is consistent with eqn.\eqref{onedim1}
only for  $p=6$, $s=1,3,5$ and $p=4$, $s=2$.
In particular,
\begin{equation}\label{onedim}
\text{\textit{in all one--dimensional representations of} $G_\mathrm{matter}$:}\quad (\bfT_s\bfL_s)^3=1.
\end{equation}

\paragraph{$V_s$ of dimension larger than 1.} If $\dim V_s>1$, both $\bfT$, $\bfL$ are
\emph{complex reflections} i.e.\! unitary matrices with all but one eigenvalues equal $1$ \cite{ST}, cfr.\! eqn.\eqref{refgr}. A group generated by complex reflections is called a \emph{complex reflection group.} We conclude that the quotient group $G_s$ of $G_\mathrm{matter}$ which acts effectively on $V_s$ is a \emph{finite}
complex reflection group.

\subsubsection{Finite complex reflection groups} \label{crg}

Finite complex reflection groups have been classified by 
Shephard and Todd \cite{ST}: there are three infinite families and 34 exceptional groups denoted $G_4,\cdots,G_{37}$. Real reflection groups (Coxeter groups) and rational reflection groups (Weyl groups) are encoded in Dynkin graphs;
complex reflection groups are also encoded in certain graphs of a more general kind \cite{Rgraph}.
We look for Shephard--Todd groups generated by two elements $\bfT$, $\bfL$ satisfying relations which imply \eqref{atleast} with $p=3,4,6$. A part for the Weyl group of $SU(2)$ (which corresponds to the already discussed $D^{(1,1)}_4$ model), there are two other  candidates in the classification list, namely the exceptional 
Shephard--Todd groups $G_4$ and $G_8$, see the following table:
\begin{table}[h]
\centering
\begin{tabular}{c|c|c|c|c|c}\hline\hline
name & graph & order & defining relations & center $Z$ & order of $Z$\\\hline
$\phantom{\bigg|}G_4$ & \begin{scriptsize}$\xymatrix{*++[o][F-]{3}\ar@{-}[r]&
*++[o][F-]{3}}$\end{scriptsize} & 24 & $\bfL^3=\bfT^3=1$, $\bfT\bfL\bfT=\bfL\bfT\bfL$ & $(\bfL\bfT)^{3k}$ & 2\\
$\phantom{\bigg|}G_8$ & \begin{scriptsize}$\xymatrix{*++[o][F-]{4}\ar@{-}[r]&
*++[o][F-]{4}}$\end{scriptsize} & 96 & $\bfL^4=\bfT^4=1$, $\bfT\bfL\bfT=\bfL\bfT\bfL$ & $(\bfL\bfT)^{3k}$ & 4\\\hline\hline
\end{tabular}
\caption{Shephard--Todd groups $G_4$ and $G_8$: graphs, orders, presentations and centers.\label{stgroups}}
\end{table}

$G_4$ is a subgroup of $U(2)$ which as an abstract group (i.e.\! forgetting its realization as a complex reflection group) is
\begin{equation}G_4\simeq \text{binary tetrahedral group $\overline{T}$}\simeq SL(2,\Z/3\Z).\end{equation}
Consider the standard Coxeter presentation
of $\overline{T}$ (which is the subgroup of $SU(2)$ corresponding to the affine graph $E_6^{(1)}$ in the McKay correspondence)
\begin{equation}\label{coxpres}\overline{T}=\big\langle A, B\;\big|\; A^2=B^3=(-AB)^3=-1\big\rangle\end{equation}
The identification
\begin{equation}\bfT=-B,\qquad \bfL=AB,\end{equation}
maps \eqref{coxpres} into the standard presentation of $G_4$. However
its explicit realization as a subgroup of $SU(2)$ does \underline{not} realize it as a complex reflection group.
To get a complex reflection group we must twist the standard quaternionic degree 2 Klein realization $Q$ of $\overline{T}$ by the character $\chi$ of one of its two non--trivial one--dimensional representations (the two choices producing equivalent results\footnote{ The two choices are related by interchanging the defining representation and its conjugate $R\leftrightarrow \overline{R}$.}).
Thus, $G_4$ is the subgroup of $U(2)$ generated by the two matrices
\begin{equation}\bfT=-\chi(B)\,B_Q,\qquad \bfL=\chi(AB)\,A_QB_Q\end{equation} 
or, explicitly,
\begin{equation}\label{defR}\bfT=-\frac{1}{\sqrt{2}}\,\omega\!\begin{pmatrix}
\epsilon & \epsilon^3\\ \epsilon & \epsilon^7\end{pmatrix},\qquad
\bfL=\frac{1}{\sqrt{2}}\,\omega\!\begin{pmatrix}
\epsilon^3 & \epsilon^5\\ \epsilon^7 & \epsilon^5\end{pmatrix}\end{equation} 
where $\omega$, $\epsilon$ are primitive roots of unity of order, respectively, $3$ and $8$. The irreducible representations of $G_4$ may be then read directly from the character table of the binary tetrahedral group $\overline{T}$.

In the same fashion, $G_8$ is a subgroup of $U(2)$ which is a central extension by $\Z/2\Z$ of the binary octahedral group $\overline{O}$, i.e.\! of the subgroup of $SU(2)$ associated to the affine graph $E^{(1)}_7$.
Explicitly \cite{ST}
\begin{equation}\label{def8}\bfT=-\frac{1}{\sqrt{2}}\,\epsilon^3\!\begin{pmatrix}
1 & -1\\ 1 & 1\end{pmatrix},\qquad
\bfL=\epsilon^3\!\begin{pmatrix}
\epsilon^3 & 0\\ 0 & \epsilon^5\end{pmatrix}\end{equation}
with $\epsilon$ as before. Again we may obtain the irreducible representations of $G_8$ by twisting the character table of $\overline{O}$.
We claim that $G_8$ is a central extension by
$Z(G_8)\equiv \Z/4\Z$ of the group $PSL(2,\Z/4\Z)$
\begin{equation}\label{claim}1\longrightarrow \Z/4\Z\longrightarrow G_8\xrightarrow{\ P\ } PSL(2,\Z/4\Z)\longrightarrow 1.\end{equation}
To show our claim, the only thing we have to prove is that
the quotient of $G_8$ by its center $Z(G_8)$
is the group $PSL(2,\Z/4\Z)$. Now
\begin{equation}\label{xx319}
G_8/Z(G_8)=\big\{\bfT, \bfL\;\big|\; \bfT\bfL\bfT=\bfL\bfT\bfL,\ (\bfT\bfL)^3=\bfT^4=\bfL^4=1\big\}\equiv PSL(2,\Z)/N(4),
\end{equation}
where $N(n)$ (for $n\in\bN$) stands for the normal closure\footnote{ $\mathsf{T}$ is the $2\times 2$ matrix defined in \eqref{2matrices}.
In eqn.\eqref{xx319} we use that $\mathsf{L}^{-1}$ is conjugate to $\mathsf{T}$ in $PSL(2,\Z)$.} of $\mathsf{T}^n$ in 
$PSL(2,\Z)$ i.e.\! the intersection of all normal subgroups of $PSL(2,\Z)$ containing $\mathsf{T}^n$. Comparing presentations, for all $n\in\bN$, one has 
\begin{equation}\begin{split}
PSL(2,\Z)/N(n)&=\text{the $(2,3,n)$ triangle group}\equiv\\
&\equiv \big\langle a, b,c \;|\: a^2=b^3=c^n=abc=1\big\rangle.
\end{split}
\end{equation}
Klein proved that $N(n)=\Gamma(n)$ for $n\leq 5$ \cite{klein}
so\footnote{ The case $n=5$ in eqn.\eqref{gggrrq} is never used in our present analysis: we listed it merely to give the complete statement of Klein's fundamental result. The case $n=6$ is radically different, since the triangle group $(2,3,6)$ yields a tessellation of the Euclidean plane, and hence is an \emph{infinite} group, while all groups of interest here must be \emph{finite} by the argument around eqn.\eqref{centralizer}.
We have already pointed out that the case of the $E^{(1,1)}_8$ model, corresponding to $p=6$, is slightly different from the other ones: using the Chinese remainder theorem,
$PSL(2,\Z/6\Z)$ is written as the product of \emph{two} triangle groups instead of a single triangle group as for $p=2,3,4$, cfr.\! eqn.\eqref{chinese}.}
\begin{equation}\label{gggrrq}
\begin{split}\Big\langle\bfT,\bfL\in &PSL(2,\Z) \;\Big|\; \bfT^n=\bfL^n=1\Big\rangle\equiv PSL(2,\Z)/N(n)=\\
&= PSL(2,\Z/n \Z)=
\left\{\begin{array}{lll} \text{dihedral} &D_5^{(1)} & n=2\\
\text{tetrahedral} & E_6^{(1)} & n=3\\
 \text{octahedral} &E_7^{(1)} & n=4\\
\text{icosahedral} &E_8^{(1)} & n=5.
\end{array}\right.
\end{split}
\end{equation}
The affine diagrams $\mathfrak{g}^{(1)}$ in the second column are the 
 McKay graphs of the \emph{double cover}
 of the $PSL(2,\Z/n\Z)$ group which is a finite subgroup of $SU(2)$. It is amusing that the three Lie algebras $E_6^{(1)}$,
$E_7^{(1)}$ and $E_8^{(1)}$ appear again in the game, now as quotients of the modular group by principal congruence subgroups $\Gamma(n)$. As abstract groups, the dihedral group of order 6 is isomorphic to $\mathfrak{S}_3$, while the tetrahedral one is isomorphic to $\mathfrak{A}_4$. 

In view of eqn.\eqref{xx319}, the case $n=4$ in \eqref{gggrrq} proves our claim \eqref{claim}.

\subsubsection{The $E^{(1,1)}_6$ model}

Let us look case by case.
For $p=3$ we have $\dim V_1=\dim V_2=3$.
Comparing the presentation of $G_4$
in table \ref{stgroups} with eqn.\eqref{atleast}
we conclude that $V_1$, $V_2$ carry a representation of $G_4\simeq SL(2,\Z/3\Z)$.
We decompose $V_s$ ($s=1,2$) into irreducible representations of $G_4$. First we rule out that $V_s$ decomposes into the direct sum of three one--dimensional representations since, in that case we would get $(\bfT\bfL)^3=1$ (by eqn.\eqref{onedim}), while we know that $(\bfT\bfL)^3$
 is the non--trivial involution  $-\pi_{12}$; its action on $V_s$ has eigenvalues $(-1,-1,1)$. We also rule out
the irreducible three dimensional representation of $G_4$ --- which coincides with the $3$--dimensional irrepresentation of $\overline{T}$ --- which, being real, cannot be a complex reflection subgroup of $U(3)$.
We remain with the defining two--dimensional representation $R$ in eqn.\eqref{defR} and its conjugate
$\overline{R}$.
For the defining representation $R$ (with the standard choice $\omega=e^{2\pi i/3})$ we have
\begin{equation}\label{wwwzq}\text{eigenvalues of $\bfT$, $\bfL$ in $R$}=(e^{-2\pi i/3},1),
\end{equation}
and we conclude that
\begin{equation}V_1=R\oplus \boldsymbol{1},\qquad
V_2=\overline{R}\oplus \boldsymbol{1},\end{equation}
where $\boldsymbol{1}$ stands for the trivial representation. As a further check, note that in the representation $R$, $(\bfT\bfL)^3=-1$ so that the eigenvalues of $(\bfT\bfL)^3$ acting on $V_s$ are $(-1,-1,+1)$ as they should.

\subsubsection{The $E^{(1,1)}_7$ model}
By comparing presentations, we see that for $p=4$ the space $V_s$ ($s=1,2,3$) carry a representation of the Shephard--Todd group $G_8$. We have
\begin{equation}\dim V_1=\dim V_3=2,\quad \dim V_2=3\end{equation}
and $(\bfT\bfL)^3$ acts on $V_s$ as $\tau^{-3}\Sigma$, i.e.\! as multiplication by
 $-e^{-3\pi i s/2}$. From eqn.\eqref{onedim}
it follows that $V_1,V_3$ are irreducible representations of $G_8$ of degree 2. There are four
such representations $F_\chi$ which are obtained by twisting the defining representation $F$ in eqn.\eqref{def8} by one of the four one--dimensional characters $\chi$ of $G_8$ (with $\chi^4=1$). One has (choosing $\epsilon=e^{\pi i/4}$)
\begin{equation}\text{eigenvalues of $\bfT$, $\bfL$ in $F_\chi$}=(\chi\, e^{3\pi i/2},\chi)
\end{equation}
so the only realizations as complex reflection groups in $\C^2$ are $F$ and its conjugate $\overline{F}$. On $F$ the central element
$(\bfT\bfL)^3$ acts as $-e^{\pi i/2}$.
Comparing with eqn.\eqref{phiact} we see that,
as representations of $G_8$,
\begin{equation}V_1=F,\qquad V_3=\overline{F}.\end{equation}

Next we consider the space $V_2$ which is spanned over $\C$ by the three strict--sense flavor charges of the QFT model.
The eigenvalues of $\bfT$, $\bfL$ in $V_2$
are $(+1,+1,-1)$ so $G_\mathrm{matter}$
acts on $V_2$ as a \emph{real}
reflection group, i.e.\! a Coxeter group.
The central element $(\bfT\bfL)^3$ acts as $1$.
Since $\bfL^2$ and $\bfT^2$ act trivially,
the group which acts effectively on $V_2$
is
\begin{equation}PSL(2,\Z)/N(2)\equiv PSL(2,\Z)/\Gamma(2)\equiv PSL(2,\Z/2\Z)\simeq\mathfrak{S}_3.\end{equation}
We have two possibilities: either $V_2$ is two copies of the trivial representation of the symmetric group $\mathfrak{S}_3$ plus the sign one--dimensional representation $\sigma$, or it is the trivial representation plus the irreducible degree 2 representation of $\mathfrak{S}_3$, $W$.
 To distinguish the two possibilities, note that in the first case $\bfT\bfL$ acts as the identity, and then (in particular)
\begin{equation}\label{diagnostics}V_2=\boldsymbol{1}\oplus\boldsymbol{1}\oplus\sigma\quad\Longrightarrow\quad\sum_{k=0}^3(-1)^k\Big\{[TLT(\tau^{-k}\co)]-[T(\tau^{-k}\co)]\Big\}=0.\end{equation}
Using eqn.\eqref{tlt1}, the above sum becomes
\begin{equation}\begin{split}
&\sum_{k=0}^3(-1)^k
\Big\{[\cs_{3,k+1}]-[\co(\vec x_3-k\vec\omega)]\Big\}=\\
&=\sum_{k=0}^3(-1)^k\Big\{[\cs_{3,k+1}]-\sum_{\ell=0}^{k}[\cs_{3,\ell}]-\sum_{i=1}^2\sum_{\ell=0}^{k-1}[\cs_{i,\ell}]\Big\}\mod\mathrm{rad}\,q
\end{split} 
\end{equation}
collecting the terms proportional to the classes localized, say, at the first special point, $[\cs_{1,\ell}]$ we see that the sum does not vanish mod $\mathrm{rad}\,q$. We conclude that
\begin{equation}V_2=W\oplus \boldsymbol{1},\end{equation}
that is: (in a suitable basis) \textit{the $3$ flavor charges 
of the $E^{(1,1)}_7$ theory are permuted by the action of the $S$--duality group} which acts through its $\mathfrak{S}_3\equiv PSL(2,\Z/2\Z)$ factor group, cfr.\! eqn.\eqref{zzz34}.
This fact was already noted in refs.\!\cite{simone}.

\subsubsection{The $E^{(1,1)}_8$ model}
For $p=6$ we have
\begin{equation}
\dim V_1=\dim V_5=1,\quad \dim V_2=\dim V_3=\dim V_4=2. 
\end{equation}
The central element $(\bfT\bfL)^3$ acts
as multiplication by $-e^{\pi i s}$ on $V_s$.

On the one--dimensional representations $V_1$, $V_5$, the generators $\bfL$, $\bfT$ act as multiplication by $e^{-\pi i/3}$ and $e^{\pi i/3}$, respectively. We denote these two characters by $\chi$ and $\overline{\chi}$.

The two conjugate representation  $V_2$ and $V_4$ are irreducible by criterion \eqref{onedim}. Acting on these representations we have
\begin{equation}\bfT^3=\bfL^3=1,\qquad \bfT\bfL\bfT=\bfL\bfT\bfL,\end{equation}
and hence the group acting effectively on $V_2$, $V_4$ is $G_4$. Acting on $V_2$
\begin{equation}
\text{eigenvalues of $\bfT$, $\bfL$}= (e^{-2\pi i/3},1).
\end{equation}
Comparing with eqn.\eqref{wwwzq}
we get
\begin{equation}V_2=R,\qquad
V_4=\overline{R}.\end{equation}

We remain with the flavor charge sublattice $V_3$ of dimension $2$.
The group acting effectively on this flavor lattice has a presentation
\begin{equation}\bfT^2=\bfL^2=1,\qquad \bfT\bfL\bfT=\bfL\bfT\bfL,\end{equation}
and hence it is identified with the Weyl group
\begin{equation}\mathrm{Weyl}(A_2)\simeq PSL(2,\Z/2\Z)\simeq \mathfrak{S}_3.\end{equation}
 Again, we have two possibilities: either \textit{(i)} $V_3$ is the direct sum of the trivial and the sign  representations, $V_3=\boldsymbol{1}\oplus\sigma$,  or \textit{(ii)}
$V_3$ is the irreducible two--dimensional representation $W$ of $\mathfrak{S}_3$. Again, the first possibility implies
$\bfT\bfL=1$. We repeat the diagnostics in eqn.\eqref{diagnostics}
\begin{equation}\label{diagnostics2}V_3=\boldsymbol{1}\oplus\sigma\quad\Longrightarrow\quad\sum_{k=0}^5(-1)^k\Big\{[TLT(\tau^{-k}\co)]-[T(\tau^{-k}\co)]\Big\}=0.\end{equation}
Explicitly the sum has the form
\begin{equation}\begin{split}
&\sum_{k=0}^5(-1)^k
\Big\{[\cs_{3,k+1}]-[\co(\vec x_3-k\vec\omega)]\Big\}=\\
&=\sum_{k=0}^5(-1)^k\Big\{[\cs_{3,k+1}]-\sum_{\ell=0}^{k}[\cs_{3,\ell}]-\sum_{i=1}^2\sum_{\ell=0}^{k-1}[\cs_{i,\ell}]\Big\}\mod\mathrm{rad}\,q
\end{split} 
\end{equation}
while the last sum does not vanish mod
$\mathrm{rad}\,q$. Again, we conclude that
\begin{equation}
V_3=W.
\end{equation}

This completes the proof of eqns.\eqref{res1}--\eqref{res3}.

\section*{Acknowledgements}
We have benefit from discussions with
Bernhard Keller and  Dirk Kussin. We thank Helmut Lenzing for
making available to us one of his unpublished manuscripts.

\appendix
\section[Explicit matrices in the canonical basis]{Explicit matrices in the canonical basis\\
and additional identities}\label{explicitmatrices}

In this appendix, we write the explicit matrices
which give the action of $T$ and $L$ on the canonical basis i.e.
\begin{equation}
[T\cl_a]=\bfT_{ab}\,\phi_b,\qquad [L\cl_a]=\bfL_{ab}\,\phi_b,
\end{equation}
where $\{\cl_a\}_{a=1}^n$ are the line bundles in eqn.\eqref{bases} such that $\phi_a=[\cl_a]$.
From section 4 we have
\begin{equation}\bfL=1- \sum_{k=0}^{p-1} 
E^t(\bphi^t)^kv\otimes v^t \bphi^k,\qquad
\bfT=1- \sum_{k=0}^{p-1} 
E^t(\bphi^t)^kw\otimes w^t \bphi^k,\end{equation}
where $E$ is the upper triangular matrix in eqn.\eqref{Eexplicit}, $\bphi=-E(E^t)^{-1}$
and
\begin{equation}v=(1,0,\dots,0)^t,\qquad w=(0,\dots,0,-1,1)^t.\end{equation}
For many purposes it is more natural to use the transpose matrices $\bfT^t$ and $\bfL^t$ giving the action on the coefficient
vectors $([X]_a)$, where, for all $X\in\mathsf{D}^b\,\mathsf{coh}\,\mathbb{X}$ we set $[X]=[X]_a\phi_a$.

The matrices $\bfT$ and $\bfL$ have quite remarkable properties, some of which were already discussed in section 4.

\subsection{$E^{(1,1)}_6$ model}\label{e6exp}

With respect to the canonical basis \eqref{bases}, one has
\begin{equation}\label{veryex3}\bfT=\left(
\begin{array}{cccccccc}
 0 & 0 & 0 & 0 & 0 & 1 & 0 & 0 \\
 -1 & 1 & 0 & 0 & 0 & 1 & 0 & 0 \\
 -1 & 0 & 1 & 0 & 0 & 1 & 0 & 0 \\
 -1 & 0 & 0 & 1 & 0 & 1 & 0 & 0 \\
 -1 & 0 & 0 & 0 & 1 & 1 & 0 & 0 \\
 0 & 0 & 0 & 0 & 0 & 0 & 1 & 0 \\
 0 & 0 & 0 & 0 & 0 & 0 & 0 & 1 \\
 -1 & 0 & 0 & 0 & 0 & 1 & 0 & 1 \\
\end{array}
\right)
\quad\bfL=\left(
\begin{array}{cccccccc}
 -1 & 1 & 0 & 1 & 0 & 1 & 0 & -1 \\
 -1 & 1 & 0 & 0 & 0 & 0 & 0 & 0 \\
 -1 & 0 & 0 & 0 & -1 & 0 & -1 & 2 \\
 -1 & 0 & 0 & 1 & 0 & 0 & 0 & 0 \\
 -1 & 0 & -1 & 0 & 0 & 0 & -1 & 2 \\
 -1 & 0 & 0 & 0 & 0 & 1 & 0 & 0 \\
 -1 & 0 & -1 & 0 & -1 & 0 & 0 & 2 \\
 -2 & 0 & -1 & 0 & -1 & 0 & -1 & 3 \\
\end{array}
\right)\end{equation}
It is easy to check that
\begin{gather}(\bfT-1)(\bfT^3-1)=(\bfL-1)(\bfL^3-1)=0,\qquad
\bfT\bfL\bfT=\bfL\bfT\bfL,\\\text{spec}\,\bfT=\text{spec}\,\bfL=(1,1,\dots,1,e^{2\pi/3}, e^{-2\pi/3}).\end{gather} 
Using the explicit matrices \eqref{veryex3}, we confirm that the central element is
\begin{equation}
(\bfT\bfL)^3=-\pi_{12},
\end{equation}
proving the claim in eqn.\eqref{Wcenter}.
Moreover
\begin{equation}(\bfT^t)^3-1=R_1\otimes M^t
\end{equation}
where $M$ is the vector of \emph{magnetic} charges of
the $\phi_a$'s, eqn.\eqref{V1}, and $R_1\equiv E^{-1}M$ is the first generator of $\mathrm{rad}\,q$, eqn.\eqref{R1},
and . Likewise
\begin{equation}(\bfL^t)^3-1=R_2\otimes Q^t
\end{equation}
where $Q$ is the vector of \emph{electric} charges of the $\phi_a$'s, eqn.\eqref{V2},
and $R_2\equiv E^{-1}Q$ is the second generator of $\mathrm{rad}\,q$. Note the similarity of the last two equations with the equation satisfied by the Coxeter element of an Euclidean algebra.

\subsection{$E^{(1,1)}_7$ model}

For the $E^{(1,1)}_7$ model the matrices are
\begin{footnotesize}
\begin{equation}\bfT=\left(
\begin{array}{ccccccccc}
 0 & 0 & 0 & 0 & 0 & 1 & 0 & 0 & 0 \\
 -1 & 1 & 0 & 0 & 0 & 1 & 0 & 0 & 0 \\
 -1 & 0 & 1 & 0 & 0 & 1 & 0 & 0 & 0 \\
 -1 & 0 & 0 & 1 & 0 & 1 & 0 & 0 & 0 \\
 -1 & 0 & 0 & 0 & 1 & 1 & 0 & 0 & 0 \\
 0 & 0 & 0 & 0 & 0 & 0 & 1 & 0 & 0 \\
 0 & 0 & 0 & 0 & 0 & 0 & 0 & 1 & 0 \\
 0 & 0 & 0 & 0 & 0 & 0 & 0 & 0 & 1 \\
 -1 & 0 & 0 & 0 & 0 & 1 & 0 & 0 & 1 \\
\end{array}
\right)\qquad
\bfL=\left(
\begin{array}{ccccccccc}
 -1 & 1 & 1 & 0 & 0 & 1 & 0 & 0 & -1 \\
 -1 & 0 & 0 & 0 & -1 & 0 & 0 & -1 & 2 \\
 -1 & 0 & 1 & 0 & 0 & 0 & 0 & 0 & 0 \\
 -1 & 0 & 0 & 0 & 0 & 0 & -1 & 0 & 1 \\
 -1 & -1 & 0 & -1 & 0 & 0 & -1 & -1 & 3 \\
 -1 & 0 & 0 & 0 & 0 & 1 & 0 & 0 & 0 \\
 -1 & 0 & 0 & -1 & 0 & 0 & 0 & 0 & 1 \\
 -1 & -1 & 0 & -1 & -1 & 0 & -1 & 0 & 3 \\
 -2 & -1 & 0 & -1 & -1 & 0 & -1 & -1 & 4 \\
\end{array}
\right)\end{equation}
\end{footnotesize}

It is easy to check that they satisfy the identities:
\begin{gather}(\bfT-1)(\bfT^4-1)=(\bfL-1)(\bfL^4-1)=0,\\
\text{spec}\,\bfT=\text{spec}\,\bfL=(1,1,\dots,1,i,-1,-i)\\
(\bfT\bfL)^3=-\bphi^{-3},\qquad\quad
\bfT\bfL\bfT=\bfL\bfT\bfL,\\
(\bfT^t)^4-1=R_1\otimes M^t,
\qquad\quad (\bfL^t)^4-1=R_2\otimes Q^t.
\end{gather}

\subsection{$E^{(1,1)}_8$ model}
For the $E^{(1,1)}_8$ model the matrices are

\begin{scriptsize}
\begin{equation}\bfT=\left(
\begin{array}{cccccccccc}
 0 & 0 & 0 & 0 & 1 & 0 & 0 & 0 & 0 & 0 \\
 -1 & 1 & 0 & 0 & 1 & 0 & 0 & 0 & 0 & 0 \\
 -1 & 0 & 1 & 0 & 1 & 0 & 0 & 0 & 0 & 0 \\
 -1 & 0 & 0 & 1 & 1 & 0 & 0 & 0 & 0 & 0 \\
 0 & 0 & 0 & 0 & 0 & 1 & 0 & 0 & 0 & 0 \\
 0 & 0 & 0 & 0 & 0 & 0 & 1 & 0 & 0 & 0 \\
 0 & 0 & 0 & 0 & 0 & 0 & 0 & 1 & 0 & 0 \\
 0 & 0 & 0 & 0 & 0 & 0 & 0 & 0 & 1 & 0 \\
 0 & 0 & 0 & 0 & 0 & 0 & 0 & 0 & 0 & 1 \\
 -1 & 0 & 0 & 0 & 1 & 0 & 0 & 0 & 0 & 1 \\
\end{array}
\right)\qquad
\bfL=\left(
\begin{array}{cccccccccc}
 -1 & 1 & 1 & 0 & 1 & 0 & 0 & 0 & 0 & -1 \\
 -1 & -1 & 0 & -1 & 0 & 0 & -1 & 0 & -1 & 3 \\
 -1 & 0 & 0 & 0 & 0 & 0 & 0 & -1 & 0 & 1 \\
 -1 & -1 & -1 & -1 & 0 & -1 & 0 & -1 & -1 & 4 \\
 -1 & 0 & 0 & 0 & 1 & 0 & 0 & 0 & 0 & 0 \\
 -1 & 0 & 0 & -1 & 0 & 0 & 0 & 0 & 0 & 1 \\
 -1 & -1 & 0 & -1 & 0 & -1 & 0 & 0 & 0 & 2 \\
 -1 & -1 & -1 & -1 & 0 & -1 & -1 & 0 & 0 & 3 \\
 -1 & -2 & -1 & -2 & 0 & -1 & -1 & -1 & 0 & 5 \\
 -2 & -2 & -1 & -2 & 0 & -1 & -1 & -1 & -1 & 6 \\
\end{array}
\right)
\end{equation}
\end{scriptsize}
which satisfy the identities:
\begin{gather}(\bfT-1)(\bfT^6-1)=(\bfL-1)(\bfL^6-1)=0,\\
\text{spec}\,\bfT=\text{spec}\,\bfL=(1,1,\dots,1,e^{\pi i/3},e^{2\pi i/3}, e^{\pi i},e^{4\pi i /3},e^{5\pi i/3})\\
(\bfT\bfL)^3=-\bphi^{-3},\qquad\quad
\bfT\bfL\bfT=\bfL\bfT\bfL,\\
(\bfT^t)^6-1=R_1\otimes M^t,
\qquad\quad (\bfL^t)^6-1=R_2\otimes Q^t.
\end{gather}

\section{Cluster--tilting}\label{tilting}

In this appendix we give some more details on the relation between the present approach to the BPS spectra of the four SCFT $D_4^{(1,1)}$, $E^{(1,1)}_6$, $E^{(1,1)}_7$, $E^{(1,1)}_8$ --- which is based on the Abelian category $\mathsf{coh}\,\mathbb{X}_p$ --- and the standard quiver approach \cite{CV11,ACCERV}
which is based on the module category of the Jacobian
algebra $\C Q/(\partial\cw)$ of the quiver $Q$ with superpotential $\cw$.

We write $T$ for the direct sum of all sheaves of the canonical basis \eqref{bases}
\begin{equation}\label{cantil}
T=\bigoplus_{\vec a\in C}\co(\vec a)\qquad C=\big\{0,\;\vec c,\; \ell_i \vec x_i, 1\leq\ell_i\leq p_i-1\big\}.
\end{equation}
We already know that the canonical algebra $\Lambda_p$ having the same weight type $(p_1,\cdots,p_s)$
as the line $\mathbb{X}_p$ is given by (here $\ch_p=\mathsf{coh}\,\mathbb{X}_p$)
\begin{equation}
\Lambda_p= \mathrm{End}_{\ch_p}(T).
\end{equation}
However, $\Lambda_p$ does not coincide with the Jacobian algebra $\C Q/(\partial\cw)$ for any choice of $(Q,\cw)$ in its mutation class.
What is true \cite{cattoy} is that there exists a $Q$ in the class which is the completion of
the quiver $Q_\mathrm{can}$ of $\Lambda_p$:
$Q$ contains $(s-2)$ extra arrows going from the sink of $Q_\mathrm{can}$ to its source and
$Q$ is endowed with a superpotential
linear in the new arrows $\eta_a$ so that
the Jacobian relations $\partial\cw/\partial \eta_a$ give back the original relations of $\Lambda_p$. For instance, for $E^{(1,1)}_r$ we add just one new arrow $\eta$ and the superpotential becomes
\begin{equation}
\cw=\eta(X_1^{p_1}+X_2^{p_2}+X_3^{p_3}).
\end{equation}

The modules of $\Lambda_p$ are then identified with the class of modules of the Jacobian algebra with $\eta=0$. One may wonder whether our treatment `forgets'
the modules with $\eta\neq0$. The answer is that these modules are already properly taken into account thanks to the properties of the cluster--category $\cc_p$.

The category $\cc_p$ has the same objects as $\mathsf{D}^b\,\ch_p$ and morphism spaces
\begin{equation}\cc_p(X,Y)=\bigoplus_{n\in\Z} \mathrm{Hom}_{\mathsf{D}^b\,\ch_p}(X,(\tau^{-1}\Sigma)^nY).\end{equation} 
and hence it is equivalent \cite{BKL} to the category $\widetilde{\ch}_p$ having the same objects as $\ch_p$ and $\Z_2$--graded morphism spaces
\begin{equation}\mathrm{Hom}_{\widetilde{\ch}_p}(X,Y)=
\mathrm{Hom}_{\ch_p}(X,Y)\oplus
\mathrm{Ext}^1_{\ch_p}(X,\tau^{-1}Y),\end{equation}
with the appropriate composition law \cite{BKL}.
The tilting object $T\in\ch_p$ in eqn.\eqref{cantil} is 
also a \textit{cluster--tilting object} \cite{clutilt}
for $\cc_p$ (for a review see \cite{reiten}).
With respect to $\ch_p$, the category $\widetilde{\ch}_p$ has additional (odd) morphisms; for instance the new morphisms $\co(\vec c)\to\co$ are
\begin{equation}\mathrm{Hom}_{\widetilde{\ch}_p}\!(\co(\vec c),\co)=
\mathrm{Ext}^1_{\ch_p}\!(\co(\vec c),\co(-\vec \omega))\simeq D\, S_{\vec c+2\vec \omega},\end{equation}
which precisely correspond to the new $(s-2)$ arrows of the completed quiver $Q$ with respect to
old $Q_\mathrm{can}$. More generally, one shows \cite{BKL,clutilt}
\begin{equation}\C Q/(\partial\cw)=\mathrm{End}_{\cc_p}\!(T)^\mathrm{op}.\end{equation} 
Then we have a functor
\begin{equation}\mathrm{Hom}_{\cc_p}(T,\cdot)\colon\cc_p\longrightarrow \mathsf{mod}\,\C Q/(\partial\cw),\end{equation}
which is full and dense; it gives an equivalence of categories
\begin{equation}\cc_p/\mathsf{add}\,\tau T\simeq \mathsf{mod}\,\C Q/(\partial\cw).\end{equation}
In particular, the indecomposable objects of
the category $\ch_p\equiv\mathsf{coh}\,\mathbb{X}_p$ are the indecomposable
modules of the Jacobian algebra together with the
$n\equiv r(\mathfrak{g})+2$ line bundles
$\co(\vec a+\vec\omega)$ with $\vec a\in C$
(cfr.\! eqn.\eqref{cantil}).

\end{document}